\documentclass{pasj00}
\usepackage{pifont}
\usepackage{multirow}
\usepackage{lscape}
\usepackage{ccaption}
\draft

\begin{document}
\SetRunningHead{N. Kudo et al.}{High Excitation Molecular Gas in the Galactic Center Loops; $^{12}$CO($J=$2--1 and $J=$3--2) Observations}

\title{High Excitation Molecular Gas in the Galactic Center Loops; $^{12}$CO($J=$2--1 and $J=$3--2) Observations}

\author{Natsuko \textsc{Kudo},\altaffilmark{1}
Kazufumi \textsc{Torii},\altaffilmark{1}
Mami \textsc{Machida},\altaffilmark{1,2}
Timothy A. \textsc{Davis},\altaffilmark{3}
Kazuki \textsc{Tsutsumi},\altaffilmark{1}
Motusuji \textsc{Fujishita},\altaffilmark{1}
Nayuta \textsc{Moribe},\altaffilmark{1}
Hiroaki \textsc{Yamamoto},\altaffilmark{1}
Takeshi \textsc{Okuda},\altaffilmark{1}
Akiko \textsc{Kawamura},\altaffilmark{1}
Norikazu \textsc{Mizuno},\altaffilmark{4}
Toshikazu \textsc{Onishi},\altaffilmark{5}
Hiroyuki \textsc{Maezawa},\altaffilmark{6}
Akira \textsc{Mizuno},\altaffilmark{6}
Kunihiko \textsc{Tanaka},\altaffilmark{7}
Nobuyuki \textsc{Yamaguchi},\altaffilmark{8}
Hajime \textsc{Ezawa},\altaffilmark{9}
Kunio \textsc{Takahashi},\altaffilmark{10}
Satoshi \textsc{Nozawa},\altaffilmark{11}
Ryoji \textsc{Matsumoto},\altaffilmark{12}
and 
Yasuo \textsc{Fukui}\altaffilmark{1}
}

\altaffiltext{1}{Department of Physics, Nagoya University, Furo-cho, Chikusa-ku, Nagoya, Aichi 464-8602, Japan}
\altaffiltext{2}{Department of Physics, Kyushu University, 6-10-1 Hakozaki, Higashi-ku, Fukuoka, Fukuoka 812-8581, Japan}
\altaffiltext{3}{Sub-Department of Astrophysics, University of Oxford, Denys Wilkinson Building, Keble Road, Oxford, OX1 3RH, UK}
\altaffiltext{4}{National Astronomical Observatory of Japan, Osawa, Mitaka, Tokyo 181-8588, Japan}
\altaffiltext{5}{Department of Physical Science, Osaka Prefecture University, Sakai, Osaka  599-8531, Japan}
\altaffiltext{6}{Solar-Terrestrial Environment Laboratory, Nagoya University, Chikusa-ku, Nagoya 464-8601, Japan}
\altaffiltext{7}{Institute of Science and Technology, Keio University, 4-14-1 Hiyoshi, Yokohama, Kanagawa 223-8522, Japan}
\altaffiltext{8}{Open Technologies Research Laboratory, 6-1-21 Hon-komagame, Bunkyo, Tokyo 113-0021, Japan}
\altaffiltext{9}{Nobeyama Radio Observatory, National Astronomical Observatory of Japan, Minamimaki, Minamisaku, Nagano 384-1305, Japan}
\altaffiltext{10}{Japan Agency for Marine-Earth Science and Technology, Kanazawa-ku, Yokohama, Kanagawa 236-0001, Japan}
\altaffiltext{11}{Department of Science, Ibaraki University, 2-1-1 Bunkyo, Mito, Ibaraki 310-8512, Japan}
\altaffiltext{12}{Faculty of Science, Chiba University, Inage-ku, Chiba 263-8522, Japan}

\email{kudo@a.phys.nagoya-u.ac.jp, fukui@a.phys.nagoya-u.ac.jp}


\maketitle

\begin{abstract}
We have carried out $^{12}$CO($J=$2--1) and $^{12}$CO($J=$3--2) observations at spatial resolutions of 1.0--3.8 pc toward the entirety of loops 1 and 2 and part of loop 3 in the Galactic center with NANTEN2 and ASTE. These new results revealed detailed distributions of the molecular gas and the line intensity ratio of the two transitions, $R_{3-2/2-1}$. In the three loops, the line intensity ratio $R_{3-2/2-1}$ is in a range from 0.1 to 2.5 with a peak at $\sim$ 0.7. The ratio in the disk molecular gas is in a range from 0.1 to 1.2 with a peak at 0.4, significantly smaller than that in the Galactic center. This supports that the loops are more highly excited than the disk molecular gas. An LVG analysis of three transitions, $^{12}$CO $J=$3--2 and 2--1 and $^{13}$CO $J=$2--1, toward six positions in loops 1 and 2 shows density and temperature are in a range 10$^{2.2}$ -- 10$^{4.7}$ cm$^{-3}$ and 15--100 K or higher, respectively. Three regions extended by 50--100 pc in the loops tend to have higher excitation conditions as characterized by $R_{3-2/2-1}$ greater than 1.2. The highest ratio of 2.5 is found in the most developed foot points between loops 1 and 2. This is interpreted that the foot points indicate strongly shocked conditions as inferred from their large linewidths of 50--100 km s$^{-1}$, confirming the suggestion by \citet{kt2009b}. The other two regions outside the foot points suggest that the molecular gas is heated up by some additional heating mechanisms possibly including magnetic reconnection. A detailed analysis of four foot points have shown a U shape, an L shape or a mirrored-L shape in the $b$-$v$ distribution. It is shown that a simple kinematical model which incorporates global rotation and expansion of the loops is able to explain these characteristic shapes. 
\end{abstract}


\section{Introduction}
The central region of the Galaxy has an outstanding concentration of molecular gas within $\sim$300 pc in radius called the Central Molecular Zone (CMZ). The molecular gas shows high temperatures and large velocity dispersions, considerably enhanced from those in the Galactic disk (e.g., \cite{mands1996}, \cite{g&p2004}). In addition, there are a few other molecular features having large velocity dispersions of 15 km s$^{-1}$ or more outside the CMZ within the central 1kpc, including Clumps 1 and 2 (\cite{tb1977}; \cite{tb1986} and \cite{as1986}) and the $l=5.5^{\circ}$ Complex (\cite{mb1997}). The origin of these high temperatures and large velocity dispersions has been a puzzle over a few decades. 

The most outstanding feature, the CMZ, harbors giant molecular clouds including the Sgr A and Sgr B2 molecular clouds (e.g., \cite{yf1977}, \cite{ssp1975}) and, in addition, the CMZ appears to be associated with two extreme velocity features called an expanding molecular ring (EMR). These features, the CMZ and EMR, form a parallelogram in a longitude-velocity diagram. Theoretical studies by \citet{jb1991} showed that the kinematics of the CMZ and EMR can be reproduced by the non-circular motion driven by a stellar bar potential. The model was applied to these features including the CMZ (e.g., \cite{hl2006}), whereas the origin of the whole anomalous kinematic features in the central kpc disk still remains unsettled (see for a review \cite{mands1996}).

It has been shown that magnetic field may also play an important role in the Galactic center as demonstrated by the Radio Arc \citep{fy1984} and non thermal filaments \citep{yhc2004}. These features suggest that the magnetic field can be as strong as mG, whereas a large volume with lower density may have weaker fields as suggested by low frequency radio observations \citep{tl2005}. A recent reanalysis of radio nonthermal radiation indicates a strong field of $\sim$50 $\mu$G over $\sim$400 pc scale \citep{cro2010}. The magnetic pressure may be important in gas dynamics since the energy density of the magnetic field is comparable to the kinetic energy density of the violent motion. Polarization measurements at sub-mm wavelength indicate that the field direction is nearly parallel to the plane \citep{gn2003}, while vertical non-thermal features indicate vertical field directions. The global field configuration yet remains to be clarified.

Fukui et al. (2006, F06) analyzed the NANTEN survey of the CO ($J=$1--0) transition toward the Galactic center \citep{mi&fu2004} and discovered two huge molecular loops in $l= 355^{\circ}$--$358^{\circ}$. These loops show enhanced velocity dispersions of 50--100 km s$^{-1}$ and F06 proposed that they are magnetically floated loops up to heights of 200--300 pc from the plane at 8.5 kpc created by the Parker instability \citep{ep1966}. F06 presented a picture that the Parker instability formed two loops and the gas falling down along the field lines accumulated as two "foot points" of molecular gas toward each end of the loops, with an estimate of magnetic field strength of 150 $\mu$G and an Alfv$\rm \acute{e}$nic speed around 24 km s$^{-1}$. Subsequently, \citet{mf2009} discovered another molecular loop, loop 3, and argued that it also represents a magnetic flotation loops. Torii et al. (2010a and 2010b, hereafter T10a and T10b) made follow up observational studies of loops 1 and 2 and a detailed analysis of muti-$J$ CO transitions in T10b derived temperature of 30--100 K or higher and density of $10^{3}$ cm$^{-3}$ in the foot points and T10b argued for shock heating in the foot points. Numerical simulations of magnetic loops were also made in a global kpc scale disk \citep{mm2009} and in a 2-D local loop \citep{kt2009}, offering a theoretical basis of the magnetic loops.

These recent studies of Galactic magnetic flotation have promoted our understanding of the properties of the loops in the Galactic center since 2006. It is important at this stage to reveal the temperature and density distributions not only toward foot points but also over the entire loops. For this purpose we have carried out new high resolution molecular observations by using the NANTEN2 4m sub-mm telescope  and the ASTE 10 m sub-mm telescope in the $^{12}$CO($J=$2--1), $^{13}$CO($J=$2--1), and $^{12}$CO($J=$3--2) transitions for loops 1 and 2 and part of loop 3. Distance is taken as 8.5 kpc (IAU 1985 standard; \cite{kl1986}). This paper is organized as follows; the observations are presented in section 2 and the results in section 3. Details of the data analysis are shown in section 4, and discussion is given in section 5. We conclude the paper in section 6.

\section{Observations}
\subsection{NANTEN 2}

Observations of the $^{12}$CO($J=$2--1) and $^{13}$CO($J=$2--1) transitions were made with the NANTEN2 4m sub-mm telescope of Nagoya University at Pampa la Bola, Chile, at an altitude of 4865 m, in February, October--November 2008 and June 2009. The half-power beam width (HPBW) was $\sim$90 arcsec at 230.538 GHz the frequency of $^{12}$CO($J=$2--1) and at 220.398681 GHz the frequency of $^{13}$CO($J=$2--1) as measured by observing the Jupiter. We observed the entirety of loops 1 and 2, and two parts of Loop 3 (the west foot point and part of the loop top) in $^{12}$CO($J=$2--1)  by using the OTF (= on the fly) mapping technique and six positions in loops 1 and 2 in $^{13}$CO($J=$2--1) using a position switching mode. 
NANTEN2 was equipped with a 230 GHz SIS receiver that provided a typical system temperature of 150--230 K in 220--230 GHz in the double-side band, including the atmosphere toward the zenith. The intensity calibration was made using the chopper-wheel method. For checking the whole system and absolute intensity calibration, M17SW [R.A.(1950) = \timeform{18h17m30s.0}, Dec.(1950) = \timeform{-16D13'6''.0}] or $\rho$--Ophiuchus [R.A.(1950) =  \timeform{16h29m20s.8}, Dec.(1950) = \timeform{-24D22'13''.5}] were observed every 2--3 hours. 

We observed Ori-KL [R.A.(1950.0) = \timeform{8h45m14s.8}, Dec.(1950.0) = \timeform{13D30'40".0} or $l$ = \timeform{213D.6695}, $b$ = \timeform{31D.7922}] and the Rosetta molecular cloud [R.A.(2000.0) = \timeform{6h34m6s.9}, Dec.(2000.0) = \timeform{4D25'37".2} or $l$ = \timeform{207D.0154}, $b$ = \timeform{-1D.8228}] with NANTEN2 for the absolute intensity calibration. We compared the spectra obtained at Ori-KL and Rosette with those obtained with KOSMA \citep{sch1998} and the 60 cm telescope of the University of Tokyo \citep{nak2007}, respectively, where all the spectra were smoothed to a beam size of 130$''$ in FWHM. We then obtained the main beam efficiency of 85 \% of NANTEN2 at 230 GHz. This efficiency was used to convert the observed value to the main beam temperature. The relative uncertainties in the calibration are estimated to be better than $\sim$5\% both in $^{12}$CO($J=$2--1) and $^{13}$CO($J=$2--1) transitions as estimated from the daily variation in the observed brightness temperature of M17SW [R.A.(1950) = \timeform{18h17m30s.0}, Dec.(1950) = \timeform{-16D13'6''.0} or $l$ = \timeform{15D.0293}, $b$ = \timeform{-0D.6722}] or $\rho$--Ophiuchus [R.A.(1950) =  \timeform{16h29m20s.8}, Dec.(1950) = \timeform{-24D22'13''.5} or $l$ = \timeform{354D.0554}, $b$ = \timeform{15D.9400}].

The spectrometer was an acousto-optical spectrometer with 2048 channels. The frequency coverage was 250 MHz, corresponding to a velocity coverage of 390 km s$^{-1}$ and a velocity resolution of 0.18 km s$^{-1}$ at 230 GHz. The telescope pointing was measured to be accurate to within 10$''$ by radio observations of the Jupiter. The final rms noise level in $T_{\rm mb}$ is $\sim$0.23 K in $^{12}$CO($J=$2--1) and $\sim$0.04 K in $^{13}$CO($J=$2--1) at 0.18 km s$^{-1}$ resolution.

\subsection{ASTE}
Observations of the $^{12}$CO($J=$3--2) line were made using the Atacama Submillimeter Telescope Experiment (ASTE) 10m sub-mm telescope of NAOJ (National Astronomical Observatory of Japan) at Pampa la Bola, Chile, at an altitude of 4865m (\cite{he2004}, \cite{he2008}) in August--September 2006 and April--June 2008. The HPBW was 22$''$ at 345.79599 GHz, the frequency of $^{12}$CO($J=$3--2). We observed the entirety of loops 1 and 2 by using a position switching mode in 2006 and an OTF mapping technique in 2008. ASTE was equipped with a 345 GHz SIS receiver, SC 345, that provided a typical system temperature of 190--300 K in 2006 and CATS 345 that provided a typical system temperature of $\sim$150 K in 2008, in the double-side band including the atmosphere toward the zenith. The beam efficiency was measured to be 60 \% for 345 GHz (ASTE web site; http://www.nro.nao.ac.jp/$\sim$aste/prop08/aste\_status.html). The intensity calibration was made using the chopper-wheel method. M17SW [R.A.(1950.0) = \timeform{18h17m30s.0}, Dec.(1950.0) = \timeform{-16D13'6''.0} or $l$ = \timeform{15D.0293}, $b$ = \timeform{-0D.6722}] or NGC6334  [R.A.(1950.0) = \timeform{17h17m32s.2}, Dec.(1950.0) = \timeform{-35D44'0''.4} or $l$ = \timeform{351D.4172}, $b$ = \timeform{-0D.6465}] was observed every 2--3 hours and the absolute intensity scale was established by comparing the intensity of M17SW between ASTE and CSO (\cite{yw1994}). The spectrometer consisted of four digital ''XF-type'' spectrometers with 2048 channels with a frequency coverage of 512 MHz, corresponding to a velocity coverage of 450 km s$^{-1}$ and a velocity resolution of 0.43 km s$^{-1}$ at 345 GHz. The telescope pointing was measured to be accurate to within 2$''$ by radio observations of the Jupiter, and W-Aql [R.A.(2000.0) = \timeform{19h15m23s.1}, Dec.(2000.0) =\timeform{-7D2'49''.8} or $l$ = \timeform{29D.7224}, $b$ = \timeform{-9D.0720}] in 2006, and RAFGL5379 [R.A.(2000.0) =  \timeform{17h44m23s.9}, Dec.(2000.0) = \timeform{-31D55'39''.1} or $l$ = \timeform{357D.6555}, $b$ = \timeform{-1D.9358}] and IRAS16594-4656 [R.A.(2000.0) =\timeform{17h03m10s.0}, Dec.(2000.0) = \timeform{-47D00'27''.6} or $l$ = \timeform{340D.7232}, $b$ = \timeform{-3D.8346}] in 2008. The final rms noise level was $\sim$0.33 K in $T_\mathrm{mb}$ in $^{12}$CO($J=$3--2) at 0.43 km s$^{-1}$ resolution.

\section{Results}
\subsection{Distributions of the molecular gas}

\begin{figure}[t]
  \begin{center}
    \FigureFile(120mm,120mm){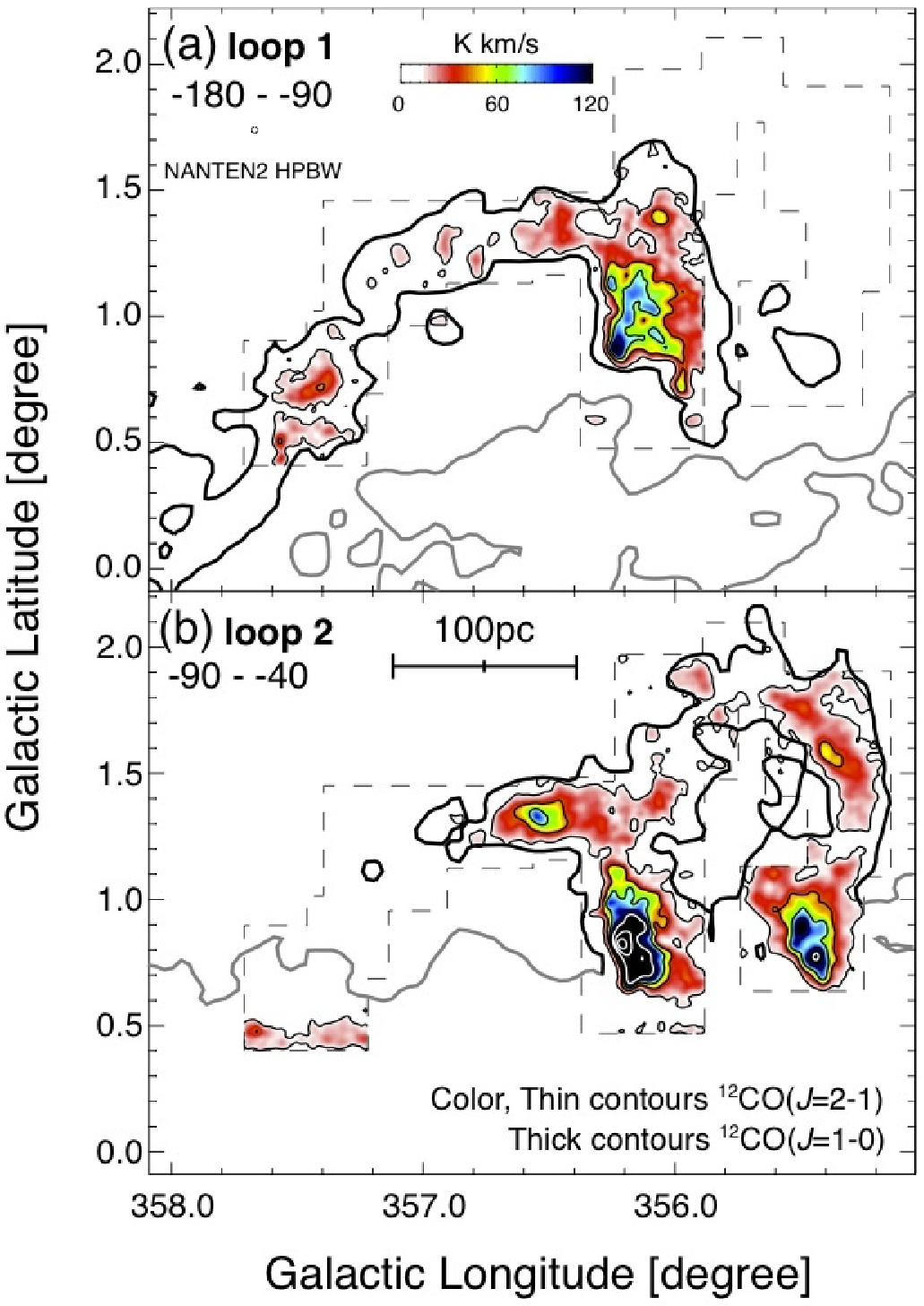}
  \end{center}
  \caption{$^{12}$CO($J=$2--1) integrated intensity distribution of loops 1 and 2. (a)Integrated intensity distribution of loop 1. The integration range in velocity is from $-180$ km s$^{-1}$ to $-90$ km s$^{-1}$. (b)Integrated intensity distribution of loop 2. The integration range in velocity is from $-90$ to $-40$ km s$^{-1}$. In both figures, color and thin contours show the $^{12}$CO($J=$1--0) emission and thick contours in black and gray show the $^{12}$CO($J=$1--0) emission. Here, thin contours in black show the loops and in gray show the other components. Thin contours are plotted every 18 K km s$^{-1}$ from 4.8 K km s$^{-1}$, and thick contours are plotted at 7 K km s$^{-1}$ and 100 K km s$^{-1}$. Dashed lines indicate the observing region of $^{12}$CO($J=$2--1).}\label{fig:CO21Loop12}
\end{figure}

\begin{figure}[t]
  \begin{center}
    \FigureFile(120mm,120mm){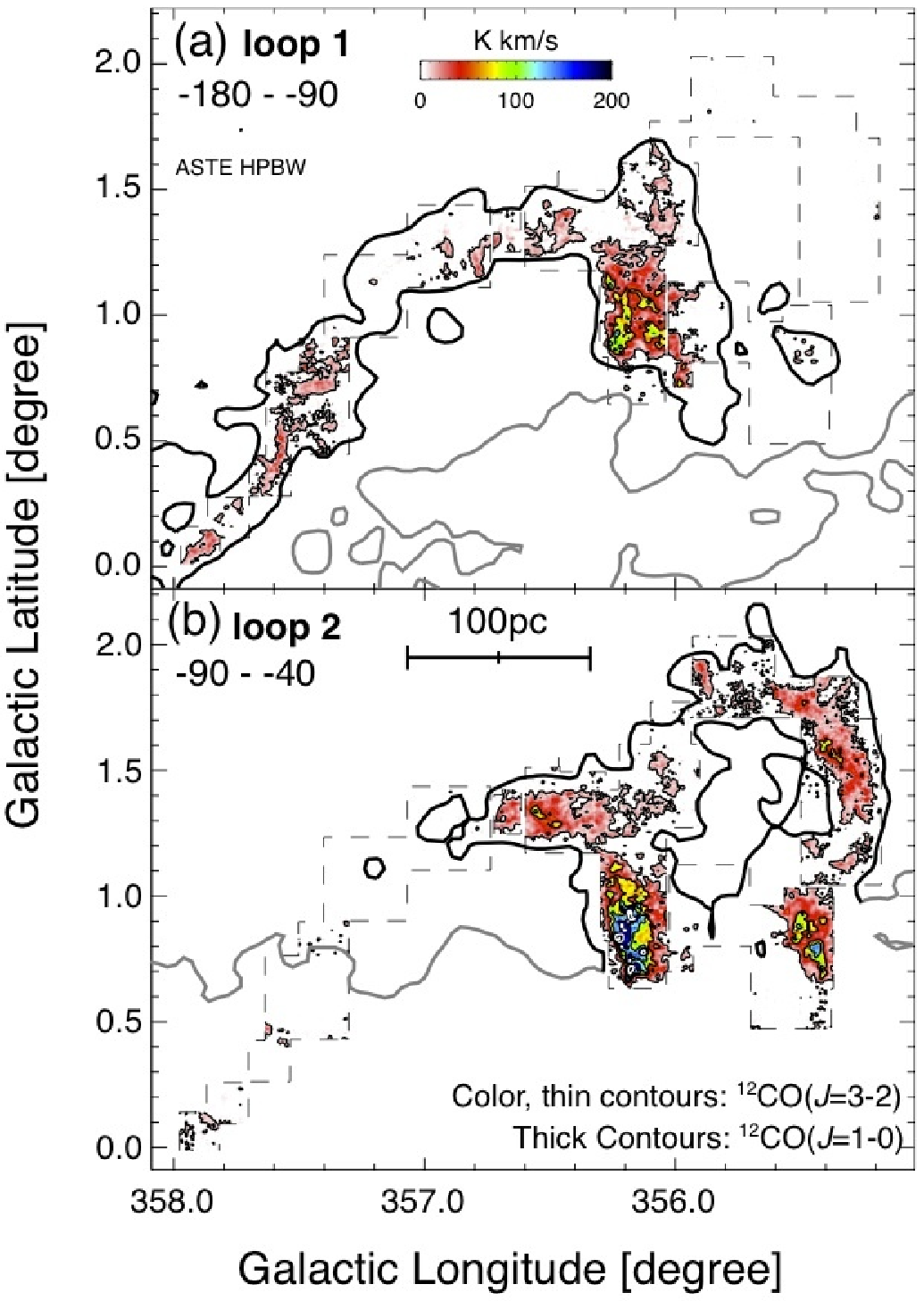}
  \end{center}
  \caption{$^{12}$CO($J=$3--2) integrated intensity distribution of loops 1 and 2. (a)Integrated intensity distribution of loop 1. The integration range in velocity is from $-180$ to $-90$ km s$^{-1}$. (b)Integrated intensity map of loop 2. The integration range in velocity is from $-90$ to $-40$ km s$^{-1}$. In both figures, color and thin contours show the $^{12}$CO($J=$1--0) emission and thick contours in black and gray show the $^{12}$CO($J=$1--0) emission. Here, thin contours in black show the loops and in gray show the other components. Thin contours are plotted every 30 K km s$^{-1}$ from 8 K km s$^{-1}$, and thick contours are plotted at 7 K km s$^{-1}$ and 100 K km s$^{-1}$. Dashed lines indicate the observing regions of $^{12}$CO($J=$3--2).}\label{fig:CO32Loop12}
\end{figure}

\begin{figure}[t]
  \begin{center}
    \FigureFile(120mm,120mm){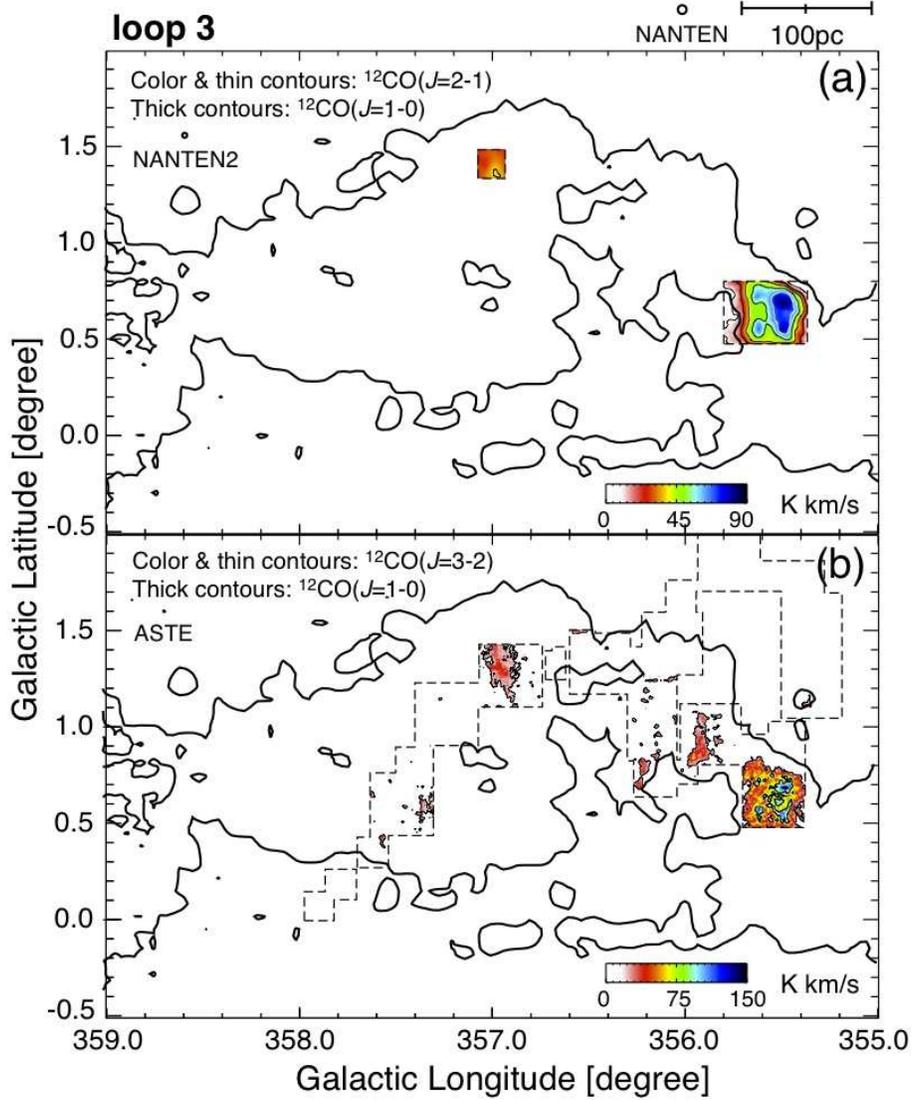}
  \end{center}
  \caption{CO integrated Intensity distributions of loop 3. The integration range in velocity is from 30 to 160 km s$^{-1}$. Color and thin contours in (a) show the $^{12}$CO($J=$2--1) emission and those in (b) the $^{12}$CO($J=$3--2) emission. The levels of the contours are the same as that in Figures \ref{fig:CO21Loop12} and \ref{fig:CO32Loop12}. In both figures, thick contours in black and gray show the $^{12}$CO($J=$1--0) emission. Here, thin contours in black show the loops and in gray show the other components. Dashed lines in (a) and (b) show the observing region of $^{12}$CO($J=$2--1) and $^{12}$CO($J=$3--2), respectively.}\label{fig:Loop3all}
\end{figure}

Figure \ref{fig:CO21Loop12} shows the integrated intensity distributions of $^{12}$CO($J=$2--1). The velocity range is $-180$--$-90$ km s$^{-1}$ including the main body of loop 1 in Figure \ref{fig:CO21Loop12}(a) and $-90$--$-40$ km s$^{-1}$ including the main body of loop 2 in Figure \ref{fig:CO21Loop12}(b) (F06). Figures \ref{fig:CO32Loop12}(a) and (b) show the intensity distributions in the same two velocity ranges in $^{12}$CO($J=$3--2). Figures \ref{fig:Loop3all}(a) and (b) show loop 3 for a velocity range 30--160 km s$^{-1}$ in the integrated intensities of $^{12}$CO($J=$2--1) and $^{12}$CO($J=$3--2), respectively. The solid thick contours in Figures \ref{fig:CO21Loop12}--\ref{fig:Loop3all} show the lowest 3$\sigma$ NANTEN $^{12}$CO($J=$1--0) contour, 7.1 K km s$^{-1}$ (T10a). 

\begin{figure}
 \begin{center}
    \FigureFile(160mm,160mm){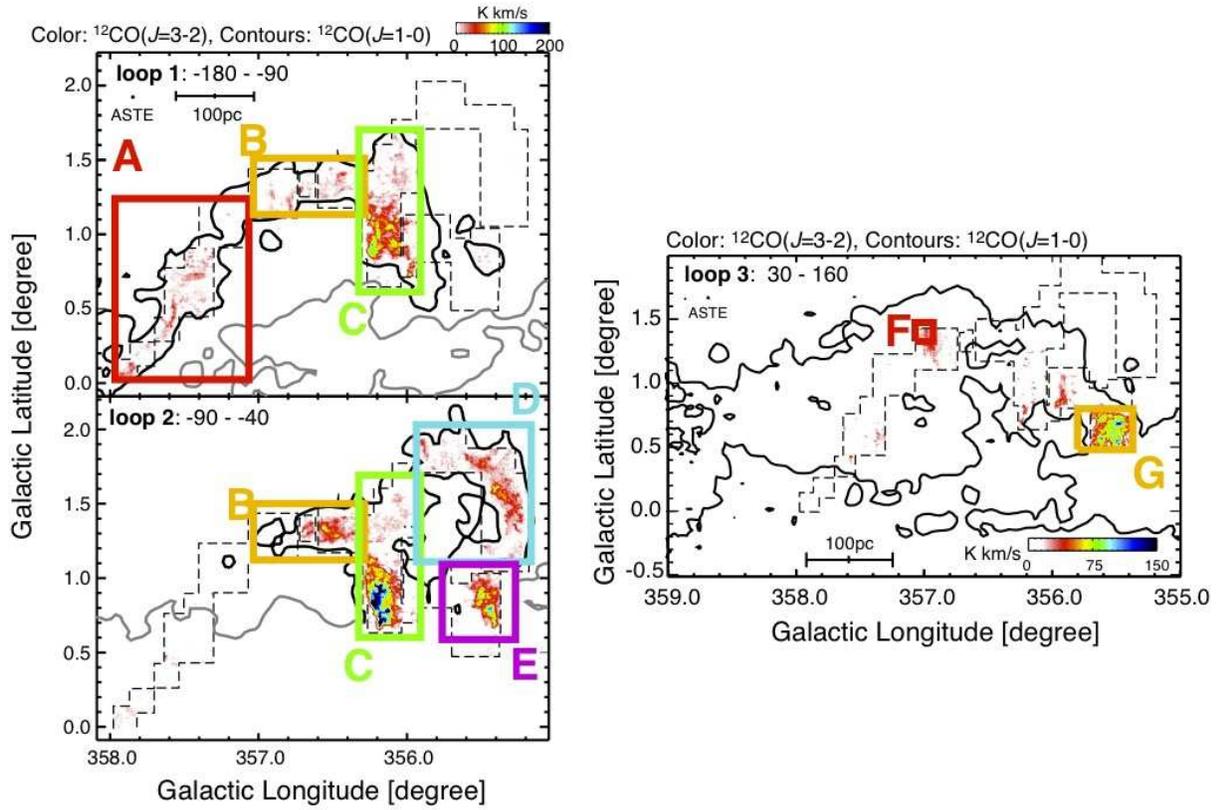}
  \end{center}
  \caption{7 regions from A to G defined in the text are shown by the boxes on the CO integrated intensity distributions of loops 1, 2 and 3. Color and contours show the integrated intensity distributions of $^{12}$CO($J=$3--2) and $^{12}$CO($J=$1--0), respectively.}
  \label{fig:CO21loop123guide}
\end{figure}

The $^{12}$CO($J=$2--1) and $^{12}$CO($J=$3--2) transitions generally show similar distributions to $^{12}$CO($J=$1--0), while the $^{12}$CO($J=$2--1) and $^{12}$CO($J=$3--2) intensities seem weaker than the $^{12}$CO($J=$1--0) intensity especially in part of the tops of loops 1 and 2. The $^{12}$CO($J=$3--2) distribution allows us to find detailed distributions at 1 pc resolution. In each panel of Figures \ref{fig:CO21Loop12} and \ref{fig:Loop3all}, the three regions of the foot points of loops 1 and 2 and the western foot point of loop 3 are clearly seen; the eastern foot point of loop 1 is located in ($l,b$)$\simeq$(\timeform{357D.2}--\timeform{357D.6}, \timeform{0D.4}--\timeform{0D.9}) (Figures \ref{fig:CO21Loop12}(a) and \ref{fig:CO32Loop12}(a)), the western foot point of loop 1 and the eastern foot point of loop 2 in ($l,b$)$\simeq$(\timeform{355D.8}--\timeform{356D.3}, \timeform{0D.6}--\timeform{1D.3}) (Figures \ref{fig:CO21Loop12}(b) and \ref{fig:CO32Loop12}(b)), the western foot point of loop 2 in ($l,b$)$\simeq$(\timeform{355D.2}--\timeform{355D.6}, \timeform{0.8}--\timeform{1D.2}) (Figures \ref{fig:CO21Loop12}(b) and \ref{fig:CO32Loop12}(b)), and the western foot point of loop 3 in ($l,b$)$\simeq$(\timeform{355D.3}--\timeform{355D.8}, \timeform{0D.5}--\timeform{0D.8}) (Figures \ref{fig:Loop3all}(a) and (b)). The present observations do not cover the eastern foot point of loop 3. We present detailed velocity channel distributions every 10 km s$^{-1}$ of the two transitions in Figures \ref{fig:channelCO10vsCO21_1} and \ref{fig:channelCO10vsCO32_1} in the Appendix 1 of Supporting Online Material. We shall denote the observed area by eight regions for discussion as shown in Figure \ref{fig:CO21loop123guide}; Regions A, B and C include loop 1, Regions C, D and E loop 2, and Regions F and G loop 3.

\subsection{Line intensity ratios}

\begin{figure}
  \begin{center}
    \FigureFile(120mm,120mm){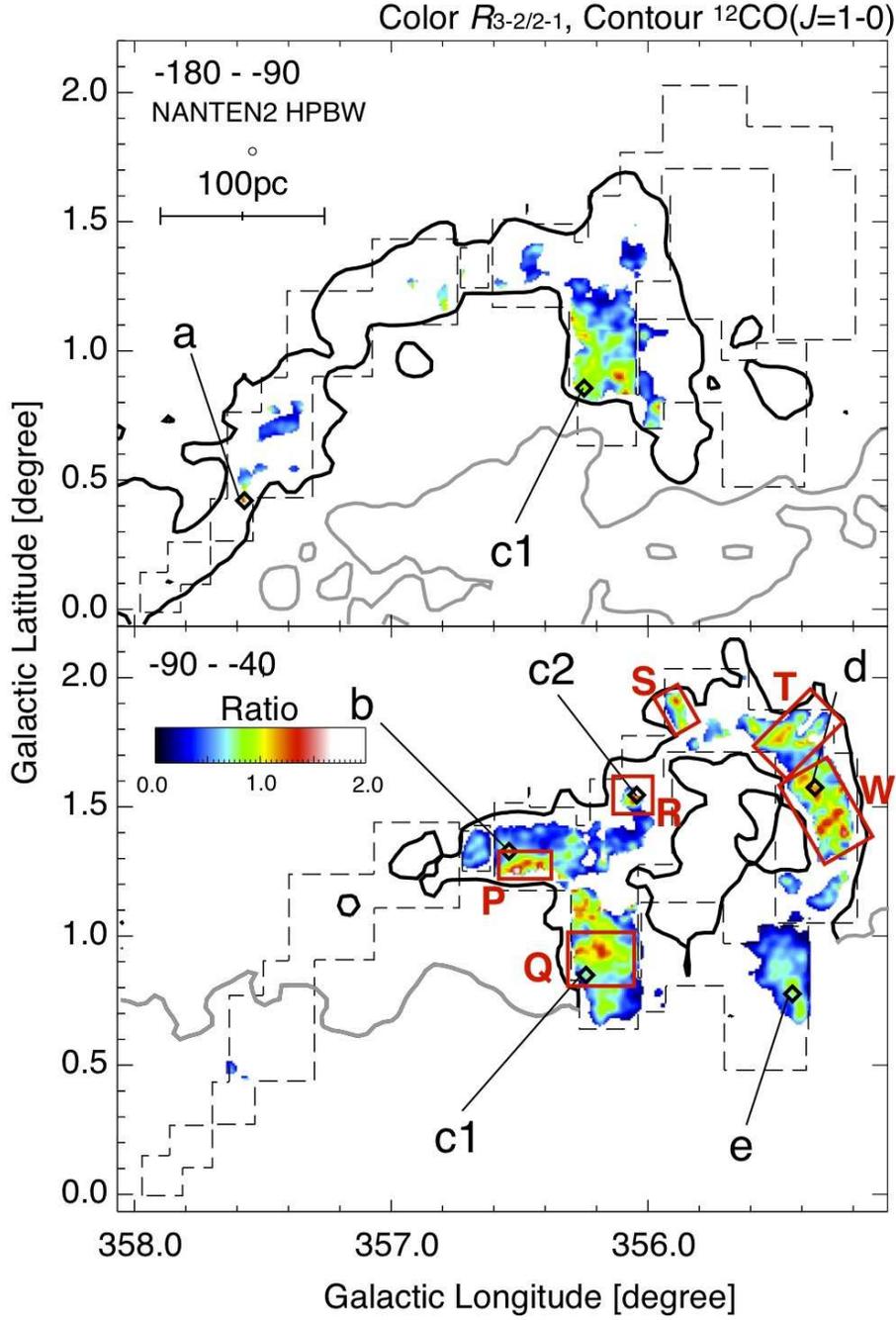}   
  \end{center}
  \caption{Distributions of the integrated intensity ratio from $^{12}$CO($J=$2--1) to $^{12}$CO($J=$3--2), $R_{3-2/2-1}$. (a)$R_{3-2/2-1}$ distributions of loop 1. The integration range in velocity is from $-180$ to $-90$ km s$^{-1}$. (b)$R_{3-2/2-1}$ distributions of loop 2. The integration range in velocity from $-90$ to $-40$ km s$^{-1}$. Contours in both figures show the $^{12}$CO($J=$1--0) integrated intensity distribution and are plotted at 7 K km s$^{-1}$. Dashed lines indicate the $^{12}$CO($J=$2--1) observing region.}
  \label{fig:ratiolb}
\end{figure}

\begin{figure}
  \begin{center}
    \FigureFile(120mm,120mm){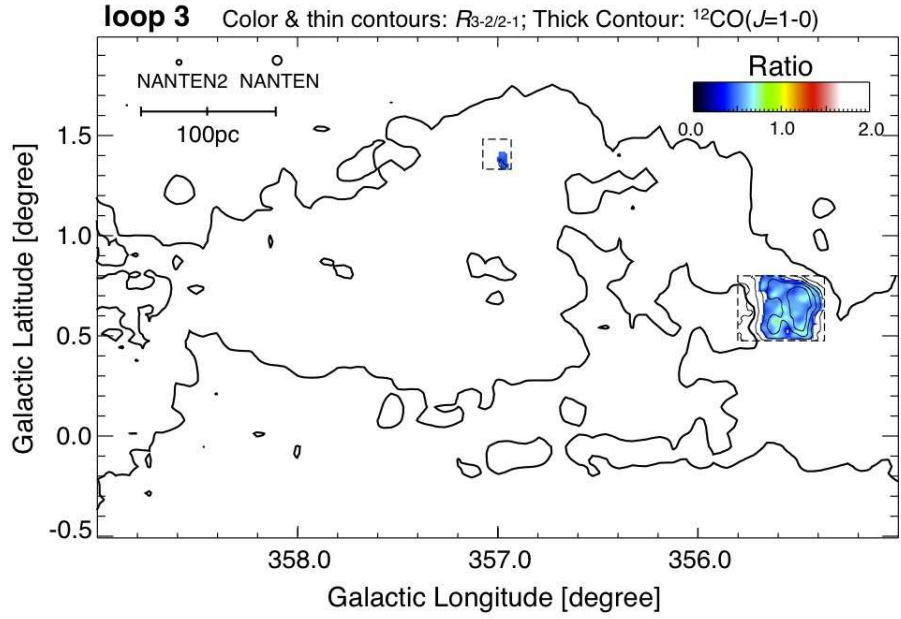}   
  \end{center}
  \caption{$R_{3-2/2-1}$ distributions of loop 3. The integration range in velocity is from $-180$ to $-90$ km s$^{-1}$. Contours show the $^{12}$CO($J=$1--0) integrated intensity distribution and are plotted at 7 K km s$^{-1}$. Dashed lines indicate the $^{12}$CO($J=$2--1) observing region.
  }
  \label{fig:ratiolbloop3}
\end{figure}

The line intensity ratio $^{12}$CO($J=$3--2)/$^{12}$CO($J=$2--1) (hereafter $R_{3-2/2-1}$) is shown for loops 1 and 2 and for loop 3 in Figures \ref{fig:ratiolb}(a) and (b), and Figure \ref{fig:ratiolbloop3}, respectively, for the same velocity range as in Figures \ref{fig:CO21Loop12} and \ref{fig:CO32Loop12}. The $^{12}$CO($J=$3--2) data is smoothed to the same Gaussian beam of 90$''$ FWHM of the $^{12}$CO($J=$2--1) data. The ratio is calculated for positions where the two line intensities are larger than the 3$\sigma$ level. Relatively high values of $R_{3-2/2-1}$ around 0.7 to 1.3 are seen in most of the loop 1 foot points in addition to the two regions toward the western inner side of the loop 1 top around ($l$,$b$)$\simeq$(\timeform{356D}.5, \timeform{1D.2}) and the western loop 2 top around ($l$,$b$)$\simeq$(\timeform{355D.3}, \timeform{1D.5}). We further note that two small regions toward ($l$,$b$)$\simeq$(\timeform{356D.0}, \timeform{1D.5}) and (\timeform{355D.9}, \timeform{1D.9}) show high values of 1.3 in $R_{3-2/2-1}$. The eastern foot point of loop 1 is not so high in $R_{3-2/2-1}$. Detailed distributions of $R_{3-2/2-1}$ every 10 km s$^{-1}$ are presented in Figure C in Appendix 1 of Supporting Online Material\footnote{The fits files of loops 1 and 2 in $^{12}$CO($J$=2--1) and $^{12}$CO($J$=3--2) can be taken from http://www.a.phys.nagoya-u.ac.jp/\~{}torii/loop/eg}. In Figure \ref{fig:ratiolb} we also indicate six areas of high line intensity ratio, P, Q, R, S, T and W as listed in Table \ref{tab:hiratiopoint}.

\begin{figure}
  \begin{center}
    \FigureFile(100mm,100mm){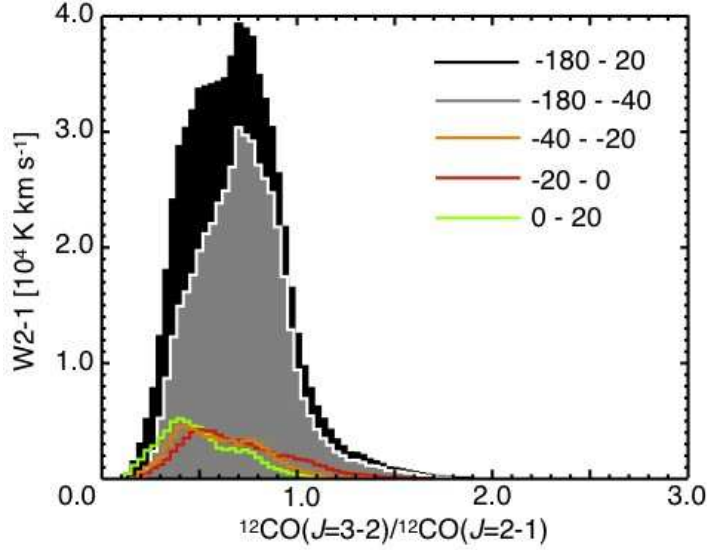}
  \end{center}
  \caption{Intensity weighted frequency distributions of $R_{3-2/2-1}$ in loops 1 and 2. The data are smoothed to a 90$''$ spatial resolution with a gaussian function and smoothed to a 2 km s$^{-1}$ velocity resolution. The intensity threshold of $^{12}$CO($J=$2--1) and $^{12}$CO($J=$3--2) is 3$\sigma$. The white lines show the contributions of the loops ($-180$ km $^{-1}$ to $-40$ km $^{-1}$), and the red, orange and green lines show the contributions of the local components with a interval of 20 km $^{-1}$ from $-40$ km $^{-1}$. The high $R_{3-2/2-1}$ region between 0 km $^{-1}$ and 15 km $^{-1}$ was subtracted from the results (see Figure C in Appendix 1 of Supporting Online Material).}
  \label{fig:histsubtract}
\end{figure}

 In Figure \ref{fig:histsubtract} we show intensity-weighted frequency distributions of $R_{3-2/2-1}$ for the velocity range of the loops, $-140$--$-40$ km s$^{-1}$, in white and the three intervals of local features in $-40$--$-20$ km s$^{-1}$ in orange, $-20$--0 km s$^{-1}$ in red, and 0--20 km s$^{-1}$ in green for the data enclosed by the broken lines in Figure \ref{fig:ratiolb}. The grey area in Figure \ref{fig:histsubtract} indicates $R_{3-2/2-1}$ in the velocity interval of the loops, where $R_{3-2/2-1}$ ranges from 0.3 to 1.2 at a 10 \% level of the maximum value at 0.7. In the three local velocity intervals from $-40$--20 km s$^{-1}$, $R_{3-2/2-1}$ histograms have peaks at 0.4--0.5.  This difference in $R_{3-2/2-1}$ suggests that the gas in the loops have more highly excited conditions than in the foreground local features, whereas the lower velocity gas may include some of the loop components. 

\begin{figure}
  \begin{center}
    \FigureFile(100mm,100mm){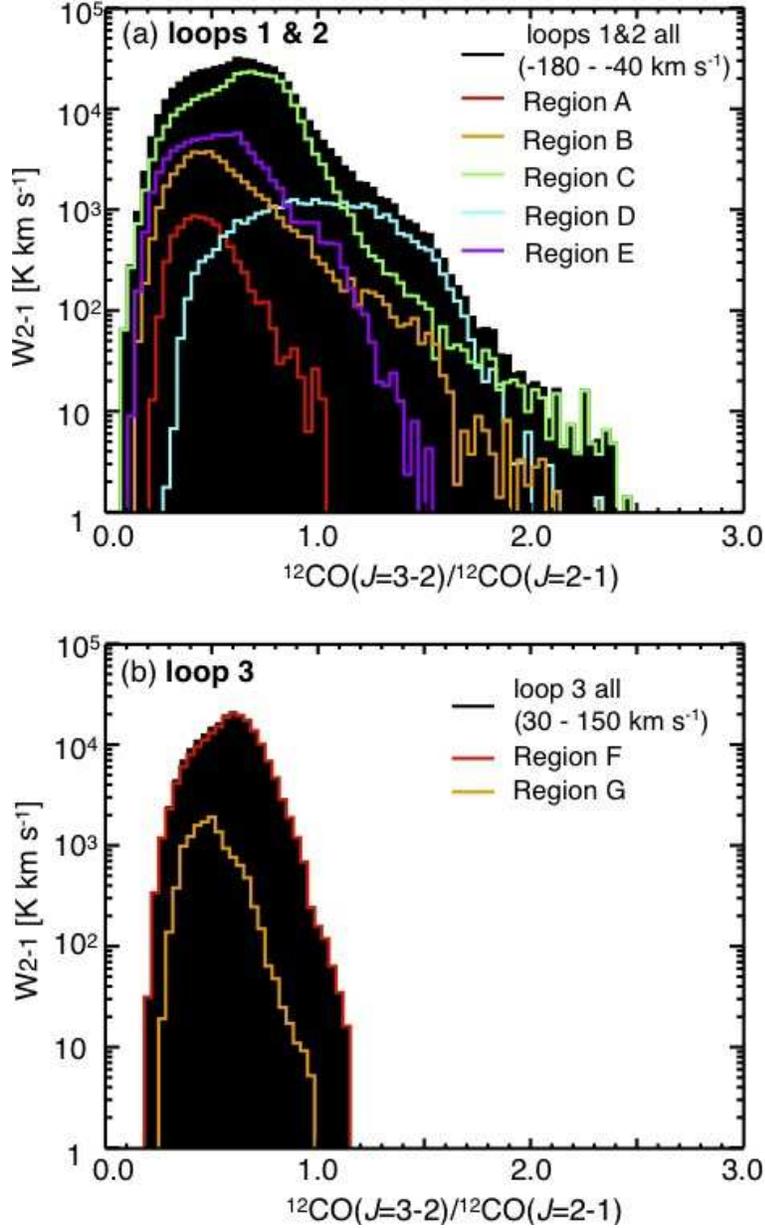}
  \end{center}
  \caption{Intensity weighted frequency distributions of $R_{3-2/2-1}$. The data are smoothed to a 90$''$  spatial resolution with a gaussian function and smoothed to a 2 km $^{-1}$ velocity resolution. The intensity threshold of $^{12}$CO($J=$2--1)and $^{12}$CO($J=$3--2) is 3$\sigma$. Colored lines in the figures show the contributions of Regions A-G shown in Figure 7.(a) $R_{3-2/2-1}$ frequency distributions of loops 1 and 2. The filled black region show the whole of loop 1 ($-140$ -- $-40$ km $^{-1}$). Red shows Region A (the east footpoint of loop 1), pink shows Region C (the right footpoint of loop 1) at $b < 1.5^{\rm \circ}$, yellow shows Region C at $b > 1.5^{\rm \circ}$, green shows Region D (the top of loop 2), and blue shows Region E (the right footpoint of loop 2). (b) $R_{3-2/2-1}$ frequency distributions of loop 3. The filled black region shows the whole of loop 3 (30 -- 160 km $^{-1}$). Red shows Region F (the loop 3 west footpoint) and orange shows Region G (the top of loop 3).}
  \label{fig:histregionseploops123}
\end{figure}

 Figure \ref{fig:histregionseploops123} shows intensity weighted frequency distributions of $R_{3-2/2-1}$ in the seven individual regions in Figure \ref{fig:CO21loop123guide}. Figure \ref{fig:histregionseploops123}(a) is for loops 1 and 2 and Figure \ref{fig:histregionseploops123}(b) for loop 3. In Figure \ref{fig:histregionseploops123}(a) Region C with the most intense foot points show a wing-like distribution reaching 2.5 in $R_{3-2/2-1}$, confirming the high excitation states found by T10b. Region B shows a similar distribution up to 2.2 and Region E up to 1.7. Regions B, C and E have peaks at 0.5--0.7 while Region D is unique in that its peak is high, around 1.3. Region A shows less values of $R_{3-2/2-1}$, whereas the observed region in $^{12}$CO($J=$2--1) in Region A does not cover the southern part of loop 1 toward ($l$,$b$)$\simeq$(\timeform{357D.5}--\timeform{358D.0}, \timeform{0D.0}--\timeform{0D.4}). The high values of $R_{3-2/2-1}$ suggest high molecular gas temperature and/or density and indicate that four Regions, B, C, D and E are in higher-excitation states.  The intensity weighted frequency distributions of $R_{3-2/2-1}$ in loop 3 are shown in Figure \ref{fig:histregionseploops123}(b). Generally, $R_{3-2/2-1}$ is less than 1.2 peaked at around 0.6 and the molecular gas in loop 3 is of lower excitation than those in loops 1 and 2.

These results indicate that Regions C and D exhibit highest excitation conditions in the observed area and Region B second highest. Region C was already analyzed by T10b. It is notable that the other foot points, Regions A and G, show lower excitation conditions and Region E has only slightly higher excitation conditions than Regions A and G (see the $R_{3-2/2-1}$ distribution in Figure \ref{fig:histregionseploops123}). So, it is not true that all the foot points show very high excitation conditions. None the less, we note that their $R_{3-2/2-1}$ distributions peaked at around 0.6 are still of higher excitation than the disk clouds that have $R_{3-2/2-1}$ peaked at around 0.4--0.5 (Figure \ref{fig:histsubtract}), and that the molecular gas in the loops show higher excitations than in the disk. 

\subsection{Details of five regions}
We describe more detailed properties of the five regions, A, C, D, E and G, in the following.\\
\underline{\textit{Region A (Loop 1 eastern foot point)}}
 
\begin{figure}
  \begin{center}
    \FigureFile(160mm,160mm){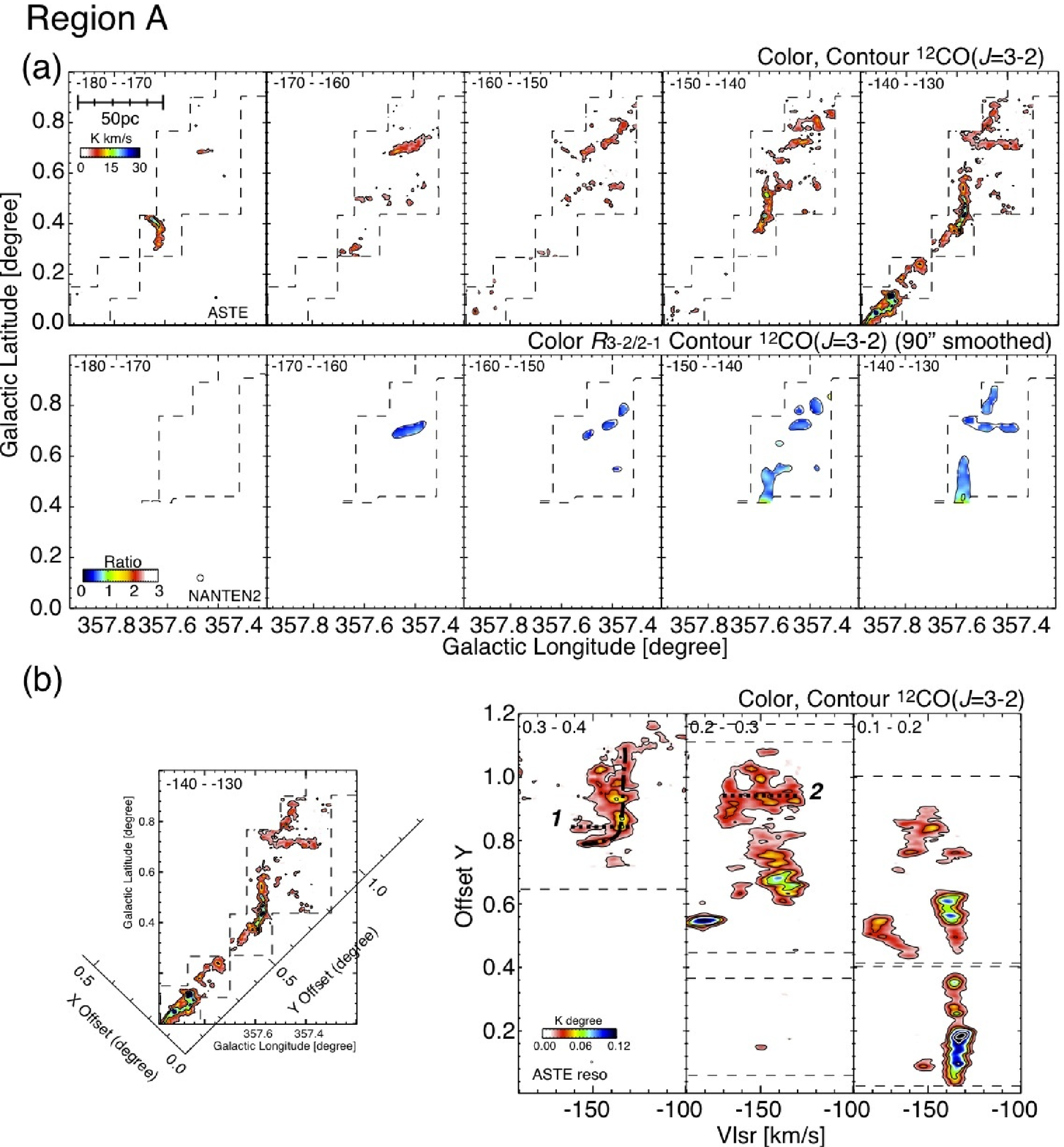}
  \end{center}
  \caption{(a-top) Velocity channel maps of $^{12}$CO($J$=3--2) integrated intensity every 10 km s$^{-1}$ toward the east footpoint of loop 1. Contours are illustrated every 10 K km s$^{-1}$ from 4 K km s$^{-1}$. (a-bottom) Velocity channel distributions of $R_{3-2/2-1}$ averaged every 10 km s$^{-1}$. Contours show the $^{12}$CO($J$=3--2) emission smoothed to a 90$''$ spatial resolution with a gaussian function and are plotted every 10 K km s$^{-1}$ from 4 K km s$^{-1}$. (b-left) Integrated intensity distribution of $^{12}$CO($J$=3--2) in Region A. An X-Y coordinate system is defined in the text. Inclination of X axis is $\sim$45$^\circ$. (b-right) X axis channel map of the $^{12}$CO($J$=3--2) in region A . Contours are plotted every 0.05 K deg. from 0.02 K deg. Hereafter, dashed-lines indicated the $^{12}$CO($J$=3--2) observing regions.}
  \label{fig:channelCO32loop1eastfp}
\end{figure}

Figure \ref{fig:channelCO32loop1eastfp}(a) shows velocity channel distributions of the $^{12}$CO($J=$3--2) intensity and $R_{3-2/2-1}$. We find filamentary distributions along the loop in $-140$--$-130$ km s$^{-1}$ in the $^{12}$CO($J=$3--2) while the $^{12}$CO($J=$2--1) observations cover only the northern part of the filament. The area around ($l$,$b$)$\simeq$(\timeform{357D.5}, \timeform{0D.8}) shows that a CO($J=$1--0) clump in T10a is resolved into a V-shaped feature and has no point whose $R_{3-2/2-1}$ is larger than 1.2.
Figure  \ref{fig:channelCO32loop1eastfp}(b) shows velocity-position distributions along the filament. We define a Y axis along the filament in an area of ($l$,$b$)$\simeq$(\timeform{357D.5}--\timeform{357D.9}, \timeform{0D.0}--\timeform{0D.4}) by taking a least-squares fitting with a straight line weighted by the integrated intensity. An X axis is taken vertical to the Y axis as shown in Figure  \ref{fig:channelCO32loop1eastfp}(b) left. Figure \ref{fig:channelCO32loop1eastfp}(b) right shows the integrated intensity every $0.1^{\circ}$ along the X axis in velocity-position distributions. Below Y = 0.4$^{\circ}$, we see only narrow features but above Y = 0.4$^{\circ}$ we find the features nos.1 and 2 has broad velocity distributions from $-180$ to $-130$ km s$^{-1}$ as indicated by dotted lines. The overall shape is like a "mirrored L" in $-160$--$-120$ km s$^{-1}$ and Y = \timeform{0.8D}--\timeform{1D.1} as indicated by dash-dotted line. We discuss these features in Section 5.2.\\
\underline{\textit{Region C (Loop 1 western foot point)}}

Figure \ref{fig:channelCO32loop1westfp}(a) shows velocity channel distributions of the $^{12}$CO($J=$3--2) intensity and $R_{3-2/2-1}$. We find two regions of high $^{12}$CO($J=$3--2) intensity; one of them in a velocity range $-100$ -- $-80$ km s$^{-1}$ in $l=356.2^{\circ}$ and $b=0.85^{\circ}$ shows a strong intensity gradient toward the plane and the other in $-60$ -- 50 km s$^{-1}$ in $l=356.2^{\circ}$ and $b=0.7^{\circ}$ shows a protrusion toward the plane. 
The gas in a velocity interval $-60$--10 km s$^{-1}$ in ($l$,$b$)$\simeq$(\timeform{356D.2}, \timeform{0D.9}) shows $R_{3-2/2-1}$ greater than 1.7 and this corresponds to the point inside the U shape discovered by T10b. This feature with high $R_{3-2/2-1}$ lead T10b to suggest that the foot point of loop 1 is extended to $-10$ km s$^{-1}$ instead of $-50$ km s$^{-1}$ as originally suggested (F06) in a $l$-$v$ distribution (Figure 2 in F06). There is a hot spot named Q in an interval $-100$--$-80$ km s$^{-1}$ in ($l$,$b$)$\simeq$(\timeform{356D.1}, \timeform{1D.5}) which shows a high $R_{3-2/2-1}$ ratio around 1.7. This feature has been discovered in the present work, since T10b does not cover $b$ greater than \timeform{1D.2}.

Figure \ref{fig:channelCO32loop1westfp}(b) shows $b$-$v$ distributions of the foot point. T10b discovered a U shape in the integration range from $l\simeq\timeform{356D.15}$ to $\timeform{356D.21}$ as indicated by dash-dotted lines and found four broad emission features, nos. 3, 4, 5 and 6, as shown by dotted lines in Figure \ref{fig:channelCO32loop1westfp}(b). Among them the feature at $l=\timeform{356D.15}$--$\timeform{356D.17}$ (no.5) shows temperature higher than 100 K (see section 4.2 of T10b). In a range from $l=\timeform{356D.15}$ to $\timeform{356D.19}$ in Figure \ref{fig:channelCO32loop1westfp}b a protrusion is indicated by bold arrows, showing a velocity dispersion of 30 km s$^{-1}$.\\
\underline{\textit{Region D (Loop 2 top)}}

Figure \ref{fig:channelCO32loop2top} shows Region D in $^{12}$CO($J=$3--2) in a velocity range $-90$--$-60$ km s$^{-1}$. We find the helical distribution noted by T10a toward ($l$,$b$)$\simeq$(\timeform{355D.72}, \timeform{1D.80}) and filamentary features, some of which are elongated normal to the loop direction toward ($l$,$b$)$\simeq$(\timeform{355D.88}, \timeform{1D.92}), (\timeform{355D.54}, \timeform{1D.80}) and (\timeform{355D.40}, \timeform{1D.76}). We find these features show higher excitation conditions toward a hot spot S in the helical distribution and toward the filamentary features in T and W in the lower panel of Figure 5.\\
\underline{\textit{Region E (Loop 2 western foot point)}}

Figure \ref{fig:channelCO32loop2westfp}(a) shows velocity channel distributions of Region E in $^{12}$CO($J=$3--2) and $R_{3-2/2-1}$, where we see a protrusion in $-90$--$-70$ km s$^{-1}$ toward ($l$,$b$)$\simeq$(\timeform{355D.4}, \timeform{0D.7}). Figure \ref{fig:channelCO32loop2westfp}(a) indicates that Region E shows $R_{3-2/2-1}$ ratio less than $\sim$1.7 and $R_{3-2/2-1}$ of this protrusion is highest in Region E. Figure \ref{fig:channelCO32loop2westfp}(b) shows a $b$-$v$ distribution of the region. We see an L-shaped feature in $l \simeq$\timeform{355D.42}--\timeform{355D.48} as indicated by a dash-dotted line. We also see three regions with broad linewidths, nos 7--9, indicated by dotted lines and the protrusion noted by bold arrows in Figure \ref{fig:channelCO32loop2westfp}(a). This distribution has a velocity dispersion of $\sim$10 km s$^{-1}$.\\
\underline{\textit{Region G (Loop 3 western foot point)}}

Figure \ref{fig:channelCO32loop3westfp}(a) shows a velocity channel distributions of Region G in $^{12}$CO($J=$3--2) and $R_{3-2/2-1}$. The gas comes closer to the left with the increase of velocity from 50 to 100 km s$^{-1}$, i.e., showing a velocity gradient of 2.0 km s$^{-1}$ pc$^{-1}$. We see two concentrations of gas toward $v\simeq$60--70 km s$^{-1}$ and ($l$,$b$)$\simeq$(\timeform{355D.5}, \timeform{0D.7}) and $v\simeq$80--90 km s$^{-1}$ and ($l$,$b$)$\simeq$(\timeform{355D.6}, \timeform{0D.6}) where $R_{3-2/2-1}$ is enhanced. Like Region A we find no gas with $R_{3-2/2-1}$ larger than 1.7. Figure \ref{fig:channelCO32loop3westfp}(b) shows a velocity-position diagram in the integration range over $l \simeq$\timeform{355D.46}--\timeform{355D.53}. We see a mirrored-L shape indicated by a dash-dotted line. We find five regions with broad linewidths, nos. 10--14, as noted by dotted lines but we do not see small protrusions extended in the north to south like that in Region C or E.

\section{Data analysis}
Figure \ref{fig:LVGplotconstdensity} shows a plot of the line intensity ratio $R_{3-2/2-1} calculated by the LVG approximation \citep{gandk1974} $ as a function of temperature of 10--200 K for density from $10^{2}$ cm$^{-3}$ to $10^{4}$ cm$^{-3}$. In these LVG calculations \citep{gandk1974}, $R_{3-2/2-1}$ ranges from $\sim$0.1 to $\sim$1.0 and is a monotonic function of density and temperature. $R_{3-2/2-1}$ becomes greater than 0.7 for density higher than $\sim10^{3}$ cm$^{-3}$ and temperature higher than $\sim$70 K. These conditions appear to characterize the high excitation gas in the Galactic center. The typical cloud condition in the Galactic disk, $10^{2}$--$10^{3}$ cm$^{-3}$ and 10--20 K, corresponds to $R_{3-2/2-1}$ less than 0.5. These properties are consistent with the trend of $R_{3-2/2-1}$ distribution in Figure \ref{fig:histsubtract}. This confirms that $R_{3-2/2-1}$ is an indicator of the excitation condition, a combination of density and temperature, while one cannot determine the parameters uniquely with these two transitions only. 

Figure \ref{fig:ratiolb} and Table \ref{tab:13COobs} show six positions where pointed observations were made in the $^{13}$CO($J=$2--1) transition with NANTEN2; they are a in Region A, b in Region B, c1 and c2 in Region C, d in Region D, and e in Region E. Figure \ref{fig:LVGspectra} shows the line profiles of the six positions, including $^{12}$CO($J=$2--1), $^{13}$CO($J=$2--1) and $^{12}$CO($J=$3--2). The $^{13}$CO($J=$2--1) emission was significantly detected in all the positions. Table \ref{tab:ratiotable} lists the ratios in the integrated intensity every 10 km s$^{-1}$ of $^{12}$CO($J=$3--2) to $^{12}$CO($J=$2--1) ($R_{3-2/2-1}$) and those of $^{13}$CO($J=$2--1) to $^{12}$CO($J=$2--1) ($R_{13/12}$). In order to estimate the physical parameters of these positions, we used an LVG analysis (\cite{gandk1974}) for the $^{12}$CO($J=$2--1), $^{12}$CO($J=$3--2) and $^{13}$CO($J=$2--1) transitions.

\subsection{LVG analysis}
In order to understand the physical conditions in molecular clouds we need to solve simultaneously the equations of line radiation transfer and statistical equilibrium of the molecular energy levels by taking into account radiative and collisional processes. In a large velocity gradient (LVG) analysis we assume a sphere with a uniform velocity gradient and uniform temperature and density. A photon emitted at any position can escape from the cloud at an escape probability $\beta = (1-\exp(-\tau))/\tau$, a function of line optical depth $\tau$, that is multiplied to the Einstein coefficients to correct for photon escaping due to the optical depth effect (\cite{jc1970}). We then estimate the line intensity of a particular transition for kinetic temperature, density and molecular abundance X(CO)/($dv/dr$), where X(CO) is the ratio of CO to H$_2$ and ($dv/dr$) the velocity gradient, by combining statistical equilibrium equations of CO rotational levels, with Einstein coefficients and collisional excitation rates (\cite{fs2005}). Here we assume that X(CO)/($dv/dr$) is $1.0 \times10^{-5}$ (T10b), [$^{12}$C]/[$^{13}$C]$\sim53$ (\cite{wr1994}; see also \cite{dr2010}) and [$^{12}$CO]/[H$_{2}$] $\sim10^{-4}$ (e.g., \cite{dl2001}).

\subsection{Results}
We have calculated the line intensities over a range of physical parameters for the line intensity values listed in Table \ref{tab:ratiotable} and calculated $\chi^{2}$ distributions between calculated ratios and observed ratios. The degree of freedom is 1 for three transitions.
Figure \ref{fig:LVGresult} shows the $\chi^{2}$ distributions of 95\% confidence level for the six positions ($\chi^{2}=3.84$) where a cross indicates the point for minimum $\chi^{2}$. Table \ref{tab:LVGtable} lists the range of physical parameters, i.e., maximum and minimum values that can be realized at 5\% probability. Figure \ref{fig:LVGresultplot} shows the results with line profiles in Table \ref{tab:LVGtable}. Typically speaking, density is $\sim10^3$ cm$^{-3}$ and temperature is in a range of 10--50 K. The positions c1 and d show highest temperatures of $\sim$100 K or even more. This is generally consistent with the distribution of $R_{3-2/2-1}$ in Figure  \ref{fig:histregionseploops123}(a) which indicates highest ratios in Regions C and D.

 The present six positions were not selected based on $R_{3-2/2-1}$ and have lower $R_{3-2/2-1}$ of 0.3--1.5, representing the whole loops, whereas Figure \ref{fig:histregionseploops123} indicates highest $R_{3-2/2-1}$ reaches 2.5. We note that the positions with $R_{3-2/2-1}$ larger than 1.0 show temperatures even higher than $\sim$ 100 K at the highest density $10^{3.5}$--$10^{4}$ cm$^{-3}$ in the LVG results (Table \ref{tab:LVGtable}). Considering these and the previous results in the loop 1 foot point by T10a, we summarize that the loops generally have temperatures around 10--50 K with highest spots of above 100 K and the typical density is $10^{3}$ cm$^{-3}$. These are significantly different high excitation conditions from those of molecular gas in the disk that has typical temperature of 10 K when no local heat source like high-mass stars is present (e.g., see \cite{am1995} for the Taurus low-mass star forming region).

\section{Discussions}
\subsection{Previous studies}
The present $^{12}$CO($J=$3--2) and $^{12}$CO($J=$2--1) study of the loops in the Galactic center has shown that the loops have higher excitation conditions of molecular transitions which are not typical of the gas in the disk. In this section we shall discuss the individual highly-excited regions in the loops. 

T10a carried out a detailed observational study of loops 1 and 2 in the CO $J=$1--0 transition at 10 pc resolution and discovered helical distributions similar to the solar helical loops, a sign of magnetic instability. T10a found counterparts to loops 1 and 2 in the H\emissiontype{I} gas and far infrared dust features and derived new estimates of the total mass and kinetic energy of the loops, $\sim 2\times10^{6}\MO$ and $\sim10^{52}$ ergs, respectively. \citet{mf2009} discovered a new loop named loop 3 in the positive-velocity range in $l\simeq\timeform{354D}$--\timeform{359D}. F06 and \citet{mf2009} argued that loops are  created by the Parker instability. More recently, T10b analyzed multiple-$J$ CO transitions including CO ($J=$1--0, $J=$3--2 and $J=$4--3) in the foot points between loops 1 and 2 and derived temperatures of 30--100 K or higher, suggesting that shock heating is responsible for raising the temperature. T10b also discovered that the foot points show a U shape in a position-velocity diagram with a few compact and broad features inside the U shape. They argued that the U shape reflects the magnetic field morphology as predicted by MHD numerical simulations and the broad features may result from magnetic reconnection between the two sides of the U shape that have anti-parallel field direction.

Theoretical studies were made to simulate formation of the magnetically floated loops. \citet{rm1988} made two dimensional magneto-hydro-dynamical numerical simulations of the process in the Galactic plane and demonstrated formation of the foot points with shock fronts. \citet{mm2009} made three dimensional simulations of a kpc-scale global nuclear disk and \citet{kt2009} made two dimensional simulations in the disk and showed that top heavy loops can be generated as is consistent with the observed molecular loops. They argued that the mechanism is similar to the formation of the solar dark filament seen in H-alpha and their Figure 10 demonstrated that the down flow of gas forms shock, heating a large volume in the foot points of a loop. 

\subsection{Physical properties of the three loops}
We note that there is no stellar heat source detected in these regions and any type of local stellar heating is not applicable here. In all these regions there are no IRAS point source candidates for protostars (\cite{tt2010}) and no radio continuum sources at 1$\sigma$ noise level (0.1 Jy beam$^{-1}$) in the radio data at 8.35 GHz in the First Galactic Plane data (GPA, \cite{gl2000}). This 8.35 GHz flux density corresponds to $10^{46}$ FUV photons s$^{-1}$, equivalent to a single B1 star of zero age main sequence (\cite{sk1994}, \cite{np1973}, \cite{pm1967}). The present LVG analysis suggests that the density is likely in a range of (1--3)$\times10^{3}$ cm$^{-3}$ and high values of $R_{3-2/2-1}$ indicate higher temperature rather than even higher density. Therefore, we require some other mechanism to explain the high excitation condition.

We summarize four possible mechanisms for the high excitation conditions already discussed in the preceding works (\cite{rm1988}, \cite{kt2009}, T10b);

\begin{description}
\item[a)] The gas in the loops falls down to the plane at Alfv\'en speed of a few 10 km s$^{-1}$ to cause shocks that compress and heat the gas (\cite{rm1988}). This type of heating is supported by a multi-line analysis of the foot points between loops 1 and 2 which derived significantly high temperature of 30--100 K or more (T10b).

\item[b)] Large-scale numerical simulations of the loops show that the gas falling down to the foot point bounces to form a spur moving upward vertical to the plane up to an altitude of several $H$ (\cite{rm1988}). The upward motion is also in the order of the Alfv$\rm \acute{e}$n speed and can cause shock heating in the spur. Such a spur is suggested in between loops 1 and 2 (T10a).

\item[c)] The magnetic reconnection where field lines become contacted in opposite directions can release thermal energy that may accelerate gas in the surroundings by pressure gradient. This is suggested to explain the broad features within the U shaped foot point in a latitude-velocity diagram in T10b.

\item[d)] The rising motion inside a loop by continuously floating gas from the plane is suggested by \citet{kt2009}. This motion can cause shock fronts inside the loop.
\end{description} 

The high temperature of foot point Q are explained by mechanism a), as shown in T10b. We shall discuss the other features below. If we assume a typical molecular clump of $2.5\times10^{4}$ $\MO$ having density $10^{3.5}$ cm$^{-3}$ and temperature 50 K, the molecular cooling rate is estimated to be $4.3\times10^{35}$ erg s$^{-1}$ (\cite{gandl1978}). Magnetic energy stored in the clump is $7\times10^{50}$ erg for 150 $\mu$G. For a hot spot like clumps in R and S with a 10 pc scale, magnetic reconnection can release energy of $10^{37}$ erg s$^{-1}$ for a typical time scale of 1 Myr. The kinetic energy of the clump is estimated to be $\sim 2\times10^{50}$ erg for motion at the Alfv\'enic speed, similar to that of the reconnection energy. We thus infer that both gravitational and reconnection energies may play a role to heat up the molecular clump by considering uncertainties in the above estimate including the unknown efficiency of kinetic energy conversion in reconnection.

Region A shows apparently weak gas concentration and a low velocity dispersion, suggesting that its shock may be weaker than in the other foot points. Its small angle to the plane, $\sim$45 degrees, may explain the weak shock. \citet{mf2009} suggested that loop 3 may be in an earlier evolutionary stage than loops 1 and 2. This leads us to suggest that Region G is not well developed yet as a foot point, explaining its weak shock. For Region E T10a noted that a mini loop-like distribution is seen above the foot point. We may speculate that the mini loop may help to reduce the falling velocity, leading to a softer shock with less excitation. To summarize, Region C shows the largest velocity dispersion and is consistent with the highest excitation conditions among all the foot points.

In the followings we shall discuss the other individual high excitation areas.
\begin{description}
\item[i)] The inside of loop 1 (P)

P is located in the south of Region B where the inner part of loop 1 is highly excited. We suggest that this local heating may be due to shock on loop 1 by mechanism d) rising gas inside the loop. On the solar surface there are a number of loops instead of a single loop (\cite{ep2004}). If we assume that loop 1 also consists of numerous unresolved loops at present, the subsequent loops rising at the Alfv$\rm \acute{e}$n speed may cause heating in the lower part of loop 1 via shocks. This is consistent with that the layer of high $R_{3-2/2-1}$ is thin. We assume that this high excitation region has size of 75 pc $\times$ 15 pc $\times$ 15 pc, and also assume density of $10^{3.5}$ cm$^{-3}$ and temperature of 50 K since $R_{3-2/2-1}$ exceeds 0.7. The cooling power of the region is $5.8 \times 10^{37}$ erg s$^{-1}$ and the cooling time scale is $\sim2.0 \times 10^{4}$ yr. This suggests that heating is very recent. We shall estimate the possible energy deposit by the rising H\emissiontype{I} gas. The H\emissiontype{I} column density is estimated by
\begin{equation}
N(\rm{H\emissiontype{I}}) = 1.82 \times 10^{18} \times \it W(\rm{H\emissiontype{I}}) \, (cm^{-2}) = 1.82 \times 10^{18} \times \it T \Delta V \, \rm{(cm^{-2})}
\end{equation}
where $W$(H\emissiontype{I})(K km s$^{-1}$) is the H\emissiontype{I} integrated intensity, $T$(K) is the main beam brightness temperature of H\emissiontype{I} and $\Delta V$(km s$^{-1}$) is the half power line width. The mass of the H\emissiontype{I} weak feature is estimated to be $M \rm(H\emissiontype{I}) \sim 1.0 \times 10^{5} \MO$ for a size of 75 pc by 75 pc, in $-180$ -- -$160$ km s$^{-1}$. If the gas is moving at the Alfv$\rm \acute{e}$n speed, 24 km s$^{-1}$, the kinetic power is estimated to be $\sim 10^{39}$ erg s$^{-1}$, much higher than the cooling energy in a given time scale. 

\item[ii)] Spur between loops (R)

R is part of the spur in Region C between loops 1 and 2 (T10b) and mechanism b) is applicable here. For a cross section of 100 pc$^{2}$ of the clump, the gas flowing upward at 10 km s$^{-1}$ has a kinetic power of $10^{36}$ erg s$^{-1}$. This is in the order of the cooling power shown above.

\item[iii)] Loop 2 tops (S, T and W)

In Region D, we see three high excitation features in $v\simeq-90$--$-60$ km s$^{-1}$; one of them is named S toward the helical distribution in $(l,b)\simeq$(\timeform{355D.88}, \timeform{1D.92}) (T10a) and the others named T and W, the filamentary distributions toward $(l,b)\simeq$(\timeform{355D.54}, \timeform{1D.80}
) and (\timeform{355D.40}, \timeform{1D.76}) (Figure \ref{fig:channelCO32loop2top}). These high excitation features are high in altitude and mechanism c) seems to be most likely. The helical distribution is able to create opposite field directions that is required for reconnection and the filamentary features in T and W may also include helical distributions yet unresolved.
\end{description} 

\subsection{Other properties}
\subsubsection{The U-shape distributions}

We presented the spatial and velocity distributions in $^{12}$CO($J=$3--2) for the four foot points in Figure \ref{fig:fplistCO32}. It is recognized that the four foot points show characteristic features in the $b$-$v$ diagrams; i.e., a U shape, or a similar L or mirrored-L shape is commonly seen in the foot points though their spatial distributions vary from one to another. These shapes are indicated by a dash-dotted line in Figures \ref{fig:channelCO32loop1eastfp}, \ref{fig:channelCO32loop1westfp}, \ref{fig:channelCO32loop2westfp} and \ref{fig:channelCO32loop3westfp} and are summarized by a schematic drawing in Figure \ref{fig:fplist+shc}. Figure \ref{fig:fplist+shc}(a) shows the three loops projected in the nuclear disk (T10a, \cite{mf2009}) and Figure \ref{fig:fplist+shc}(b) a $b$-$v$ diagram of the four foot points. The two foot points nearer to the sun, Regions A and G, have concentration of gas on the larger velocity side and part of the gas in the bottom of the U shape tends to extend to the lower velocity side. On the other hand, the gas in the two foot points, Regions C and E, is concentrated on the lower velocity side. Their apparent velocity extents are $\sim$ 20 km s$^{-1}$ for Region A, $\sim$60--80 km s$^{-1}$ for Region C, $\sim$35 km s$^{-1}$ for Region E, and $\sim$30 km s$^{-1}$ for Region G. It is notable that Region C shows the largest velocity dispersion, whereas the other three have smaller velocity dispersions of 20--35 km s$^{-1}$. 

The gas in Region C in general shows high values in $R_{3-2/2-1}$ over 1.0 (Figure \ref{fig:channelCO32loop1westfp}a). The U-shaped distribution was discovered by T10b and these authors suggested that the U shape may be formed in the magnetic flotation by the down flowing gas on both sides of a loop. This interpretation is qualitatively applicable to all the foot points. The velocity dispersions of 20--35 km s$^{-1}$ of the three foot points except for Region C are consistent with the Alfv$\rm \acute{e}$n speed 24 km s$^{-1}$ estimated in F06. Quantitatively, however, Region C shows a much larger velocity dispersion and requires some additional velocity separation as noted by T10b. The foot points of loops 1 and 2 for instance may be separated in space, causing a larger velocity separation due to different rotation and/or expansion velocity.

We here attempt to model the basic features in the foot points by adopting the expansion and rotation derived by T10a and the falling motion of 2-D numerical simulations (\cite{kt2009}). Details are given in Appendix 2 of Supporting Online Material. The model indicates that the kinematics is basically dominated by the expansion and rotation of 50--150 km s$^{-1}$ and that the foot point shows U or L/mirrored-L shapes as observed in $b$-$v$ diagrams (Figure \ref{fig:fplist+shc}). The model therefore qualitatively reproduces the observations, whereas qualitative details like velocity spans are not well fit. This is actually expected because the 2-D model is not finely tuned to fit the observations. For instance the latitude of the foot points is too low as compared with the observations and the velocity widths are not taken into account. We also note that assumptions adopted in T10a and \citet{mf2009} are too simple, allowing only a first-order approximation at best. Nevertheless, we regard the basic shapes like U and L/mirrored-L support that the idea of the model is in the right direction.

\subsubsection{Compact broad features}
We shall discuss the broad features that appear to connect two velocity features in U shapes of the foot points (Figures \ref{fig:channelCO32loop1eastfp}, \ref{fig:channelCO32loop1westfp}, \ref{fig:channelCO32loop2westfp} and \ref{fig:channelCO32loop3westfp} shown by dotted lines). We numbered them as 1--14 and list their integrated intensity and line intensity ratios in Table \ref{tab:cloudtable}. T10b noted one of the features no.5 in Region C for the first time and more than ten similar features indicate that the compact broad features are common in the foot points.

All the regions with high excitation conditions do not have heating sources (See Section 5.2). The total gas mass with large velocity dispersions in Region C is estimated to be $\sim$ $1.0 \times 10^{3}$ $\rm \MO$ for no.5 (T10b, by using the Mopra $^{12}$CO($J=$1--0) data with an X factor of $\sim 0.7\times10^{20}$ (cm$^{-2}$/(K km s$^{-1}$), T10a) and the total kinetic energy is estimated to be $4 \times 10^{48}$ erg for velocity dispersion $\sim20$ km s$^{-1}$. Magnetic reconnection between field lines in the opposite directions may have occured in the U shape as discussed by T10b. Assuming reconnection is working, we shall compare the energy of the magnetic field and the kinetic energy of the gas with a large velocity dispersions in Region C. The total magnetic energy is estimated to be $4.0 \times 10^{50}$ erg (T10b). If the magnetic energy is converted into this kinetic energy, the conversion efficiency is 2 \%, indicating that the magnetic reconnection is a viable explanation. 

\subsubsection{Protrusions}
Figures \ref{fig:channelCO32loop1westfp} and \ref{fig:channelCO32loop2westfp} show Regions C and E have protrusions as indicated by a bold arrow. These have velocity spans of 20--40 km s$^{-1}$ over 20 pc and are elongated nearly vertical to the plane. In the protrusions, the ratio $R_{3-2/2-1}$ is around $\sim$0.7, higher than that in the surroundings of $\sim$0.4. We speculate that the origin of these protrusions is due to downward force either by magnetic reconnection or bouncing in the foot point. These are intriguing features which may characterize the foot points, whereas their energy is not dominant. Such features are not predicted by numerical simulations (\cite{kt2009}) and more realistic simulations are desirable to pursue these details.

\section{Conclusions}
We have carried out $^{12}$CO($J=$2--1) and $^{12}$CO($J=$3--2) observations at spatial resolutions of 1.0--3.8 pc toward the entirety of loops 1 and 2 and part of loop 3 in the Galactic center with NANTEN2 and ASTE. These new results revealed detailed distributions of the molecular gas and the line intensity ratio of the two transitions, $R_{3-2/2-1}$. Main points of the study are summarized below;

\begin{enumerate}
\item The $^{12}$CO $J=$3--2 and 2--1 transitions show similar distributions to $^{12}$CO $J=$1--0 in loops 1 and 2. In the three loops, the line intensity ratio $R_{3-2/2-1}$ is in a range from 0.1 to 2.5 with a peak at $\sim$0.7. The ratio in the disk molecular gas is in a range from 0.1 to 1.2 with a peak at 0.4, significantly smaller than that in the Galactic center. This suggests that the loops are more highly excited than the disk molecular gas. 

\item An LVG analysis of three transitions, $^{12}$CO $J=$3-2 and 2--1 and $^{13}$CO $J=$2--1, toward six positions in loops 1 and 2 shows that the density and temperature are in a range $10^{2.2}$--$10^{4.7}$ cm$^{-3}$ and 15--100 K or higher, respectively. Considering the frequency distribution of $R_{3-2/2-1}$, we suggest that the loops are characterized by higher temperature and/or density significantly different from those in the Galactic disk molecular gas. 

\item We have found that three regions extended by 50--100 pc in the loops tend to have higher excitation conditions as characterized by  $R_{3-2/2-1}$ greater than 1.2. These include a region inside loop 1(Region B), a region in the foot points of loops 1 and 2 (Region C) and a region above the foot point of loop 2 (Region D). In addition, we find two hot compact spots with high  $R_{3-2/2-1}$ above Region C. The highest ratio of 2.5 is found in the most developed foot points in Region C. The high ratio in the foot points is interpreted as evidence for strongly shocked conditions which is consistent with their large linewidths of 50--100 km s$^{-1}$. The other two regions outside the foot points suggest that the molecular gas is heated up by some additional heating mechanism other than the gravitational energy release possibly including magnetic reconnection, since the regions have no associated young massive stars.

\item A detailed analysis of four of the foot points have shown a U shape, an L shape or a mirrored-L shape in the latitude-velocity distribution. These shapes share similarity with the U shape reported in the foot points between loops 1 and 2 by T10b. The U and L/mirrored-L shaped features have large velocity widths of 20--100 km s$^{-1}$ and have broad features in their bottom. It is shown that a simple kinematical model which incorporates global rotation and expansion of the loops in addition to falling motion along the loops is able to explain the characteristic shapes. 

\item We found that the foot points have localized broad features that appear to lie between the two velocity features constituting the U shapes. Because these broad features have no associated young stars, they require some non stellar energy source, where the typical kinetic energy involved in one of such broad features is $10^{49}$ erg. We argue that the broad compact features may be due to magnetic reconnection in the U shaped gas where magnetic field lines in the opposite directions become reconnected. We present also found protrusions in the bottom of foot points possibly formed by magnetic reconnection or bouncing in the foot points.
\end{enumerate}

\bigskip
We thank the all members of the NANTEN2 consortium and ASTE team for the operation and persistent efforts to improve the telescopes. The NANTEN2 project is an international collaboration between 10 universities: Nagoya University, Osaka Prefecture University,  University of Cologne, University of Bonn, Seoul National University, University of Chile, University of New South Wales, Macquarie University, University of Sydney and Zurich Technical University. The ASTE project is driven by Nobeyama Radio Observatory (NRO), a division of National Astronomical Observatory of Japan (NAOJ), in collaboration with University of Chile, and Japanese institutes including University of Tokyo, Nagoya University, Osaka Prefecture University, Ibaraki University, and Hokkaido University. Observations with ASTE were in part carried out remotely from Japan by using NTT's GEMnet2 and its partner R\&E (Research and Education) networks, which are based on AccessNova collaboration of University of Chile, NTT Laboratories, and NAOJ. NANTE2 project is based on a mutual agreement between Nagoya University and the University of Chile and includes member universities, Nagoya, Osaka Prefecture, Cologne, Bonn, Seoul National, Chile, New South Wales, Macquarie, Sydney and Zurich. 
This research was supported by the Grant-in-Aid for Nagoya University Global COE Program, ''Quest for Fundamental Principles in the Universe: from Particles to the Solar System and the Cosmos'', from the Ministry of Education, Culture, Sports, Science and Technology of Japan. This work is financially supported in part by a Grant-in-Aid for Scientific Research
(KAKENHI) from the Ministry of Education, Culture, Sports, Science and Technology of Japan (Nos. 18026004 and 21253003) and from JSPS (Nos. 14102003, 20244014, and 18684003) and from the Sumitomo Foundation and the Mitsubishi Foundation. This work is also financially supported in part by core-to-core program of a Grant-in-Aid for Scientific Research from the Ministry of Education, Culture, Sports, Science and Technology of Japan (No. 17004).

\clearpage

\appendix
\section{Velocity channel distributions of $^{12}$CO($J=$2--1), $^{12}$CO($J=$3--2) and $R_{3-2/2-1}$}
Figures \ref{fig:channelCO10vsCO21_1} and \ref{fig:channelCO10vsCO32_1} show velocity channel distributions of $^{12}$CO($J=$2--1) and $^{12}$CO($J=$3--2) in $l$ and $b$, respectively\footnote{The fits files of loops 1 and 2 in $^{12}$CO($J$=2--1) and $^{12}$CO($J$=3--2) can be taken from http://www.a.phys.nagoya-u.ac.jp/\~{}torii/loop/eg}. Loop 1 is for a velocity range from $-160$ km s$^{-1}$ to $-50$ km s$^{-1}$, loop 2 from $-120$ km s$^{-1}$ to $-30$ km s$^{-1}$ and loop 3 from 0 km s$^{-1}$ to 120 km s$^{-1}$. The thick contours show the lowest 3$\sigma$ level of the $^{12}$CO($J=$1--0) intensity, 5.0 K km s$^{-1}$. It is not shown in the present paper but we note that $^{12}$CO($J=$3--2) emission is detected in loop 3 top in $(l,b)\simeq$(\timeform{357D.5}--\timeform{356D.8}, \timeform{0D.3}--\timeform{1D.7}) and $v=$60--120 km s$^{-1}$. Loops 1 and 2 tend to become separated from the center with the increase of velocity while loop 3 shows an opposite trend. \citet{mf2009}, and T10a give a simple kinematical model to explain this velocity distribution by assuming rotational motion with expansion. 

The local features outside the Galactic center have lower velocities of $-40$--20 km s$^{-1}$ and they show different distributions in $^{12}$CO($J=$1--0), $^{12}$CO($J=$2--1) and $^{12}$CO($J=$3--2). The $b$ distribution of the $^{12}$CO($J=$1--0) and $^{12}$CO($J=$2--1) is extended to $b=$1.0$^{\circ}$, whereas the $^{12}$CO($J=$3--2) is limited to $b$ less than \timeform{0D.25} in 0--20 km s$^{-1}$. This difference suggests that the $^{12}$CO($J=$3--2) selectively traces denser/warmer gas near the plane than the lower $J$ lines and we may separate the loop features from the foreground local features at higher angular resolutions.

Figure \ref{fig:channelratioCO21vsCO32_1} shows velocity channel distributions of the line intensity ratio $R_{3-2/2-1}$ for loops 1 and 2. The regions where $R_{3-2/2-1}$ is greater than 1.5 are four regions; 1) toward the loop 1 top in $v\simeq-80$--$-70$ km s$^{-1}$ and $(l,b)\simeq$(\timeform{356D.5}, \timeform{1D.2}), 2) toward the loop 2 top in $v\simeq-60$--$-50$ km s$^{-1}$ and $(l,b)\simeq$(\timeform{355D.3}, \timeform{1D.5}), 3) toward the loop 1 west top in $(l,b)\simeq$(\timeform{356D.2}, \timeform{0D.9}), and 4) in $v\simeq0$--10 km s$^{-1}$ around $(l,b)\simeq$(\timeform{355D.2}, \timeform{1D.5}). The fourth region has very small linewidths of a few km s$^{-1}$ in $^{12}$CO($J=$3--2) in the velocity range of local features and we shall not deal with this in the present work.

\section{Modeling the velocity distribution of the foot points}

We make a simple kinematic modeling of the three loops.  

Figure \ref{fig:loopmodel12} presents the model for loops 1 and 2 where we adopt a radius of the loops 670 pc with an expansion velocity of 150 km s$^{-1}$ and a rotational velocity of 50 km s$^{-1}$ (Section 6.3 of T10a).  The position angle of $55^{\circ}$ is adopted for the position angle between loops (Figures \ref{fig:loopmodel12}(a) and \ref{fig:loopmodel12}(b)). The falling velocity and the loop shape are taken from the numerical simulation by \citet{kt2009} at an elapsed time of $5.25\times10^{7}$ yr so as to fit the loop height (Figure \ref{fig:loopmodel12}(b)). A $l$-$b$ diagram and a $b$-$v$ diagram are shown in Figures \ref{fig:loopmodel12}(d) and \ref{fig:loopmodel12}(e). Figure \ref{fig:loopmodel3} is for loop 3 whose radius is 1000 pc (Section 5.4 of \citet{mf2009}) with a position angle of the center of the loop, $145^{\circ}$ (Figures \ref{fig:loopmodel3}(a) and \ref{fig:loopmodel3}(b)). The expansion velocity is 130 km s$^{-1}$ and the rotational velocity is 80 km s$^{-1}$. The falling velocity and the shape is tentatively taken from a model at $7.84\times10^{7}$ yr from \citet{kt2009} to fit the loop height (Figure \ref{fig:loopmodel3}(b)). An $l$-$b$ diagram and a $b$-$v$ diagram are shown in Figures Ed and \ref{fig:loopmodel3}(e). The dotted lines in Figures \ref{fig:loopmodel12} and \ref{fig:loopmodel3}(e) indicate that without falling motion.

These model calculations show that the foot points show U or L/mirrored-L shaped distributions in $b$-$v$ diagrams, independently of the falling motion. This indicates that the U and L/mirrored-L shapes basically reflect expansion and rotation. Figures \ref{fig:loopmodel12} and \ref{fig:loopmodel3}(e) show boxes of Regions A, C, E and G. The trend of U and L/mirrored-L shapes is qualitatively reproduced as is consistent with the observations. We, however, note that the quantitative agreement is not currently good. This is perhaps because the model by \citet{kt2009} has no fine tuning to fit the observations and, rather, intended to reproduce physical processes on the loops.

\newpage

\begin{table}
  \caption{Positions of hot regions in -90 - -40 km s$^ {-1}$}\label{tab:hiratiopoint}
  \begin{center}
    \begin{tabular}{ccccc}
\hline
\hline
ID & $l$($^{\circ}$)\footnotemark[$*$] & $b$($^{\circ}$)\footnotemark[$*$] & max $R_{3-2/2-1}$ & reference\footnotemark[$\dagger$]\\
\hline 
P & 356.40 	& 1.24 	& 1.90 	& Region B\\
Q & 356.28 	& 0.92 	& 1.28 	& Region C\\
R & 356.04 	& 1.54 	& 1.55 	& Region C\\
S & 355.89 	& 1.90 	&  1.15 	& Region D\\
T & 355.37 	& 1.76 	&  1.27 	& Region D\\
W & 355.23 	& 1.38 	& 1.53 	& Region D\\
\hline
\multicolumn{4}{@{}l@{}}{\hbox to 0pt{\parbox{160mm}{
       \footnotemark[$*$] These positions indicate maximum $R_{3-2/2-1}$ positions.
      \par\noindent
      \footnotemark[$\dagger$]  These regions are shown in Figure 4.
       \par\noindent
   }\hss}}
    \end{tabular}
  \end{center}
\end{table}

\begin{table}
  \caption{Positions of $^{13}$CO($J=2-1$) observations}\label{tab:13COobs}
  \begin{center}
    \begin{tabular}{ccccccc}
\hline
\hline
\multirow{2}{*}{ID} & $l$ & $b$ & \multicolumn{4}{c}{$^{12}$CO($J=2-1$)} \\
           \cline{4-7}
          & ($^{\circ}$) & ($^{\circ}$) & T$_{peak}$(K) & V$_{peak}$(km s$^{-1}$) & $\Delta$V(km s$^{-1}$) & Hot regions\\
\hline 
a & 357.58 	& 0.42	& 1.7 	& $-139.5$ & 14.0&\\
b & 356.55 	& 1.33	& 3.2 	& $-67.3$ & 39.7& P\\
c1 & 356.25 	& 0.84 	& 7.0 	& $-90.0$ & 58.0 &Q\\
c2 & 356.06 	& 1.52 	& 0.8 	& $-94.1$ & 48.3&R\\
d & 355.38 	& 1.57 	& 1.9 	& $-62.0$ & 26.5&U\\
e & 355.53 	& 0.80 	& 3.9 	& $-75.7$ & 53.5\\
\hline
    \end{tabular}
  \end{center}
\end{table}

\begin{table}[p]
  \caption{$R_{3-2/2-1}$ and $R_{13/12}$ in peaks a--e}\label{tab:ratiotable}
  \begin{center}
    \begin{tabular}{cccc}
\hline
\hline
\multirow{2}{*}{Peak} & V$_{\rm LSR}$ range & \multicolumn{2}{c}{Line ratio}\footnotemark[$*$]  \\
           \cline{3-4}
           & (km s$^{-1}$) & $R_{3-2/2-1}$ &  $R_{13/12}$ \\
\hline 
            \multirow{6}{*}{a} 
            &	$-$155--$-145$	& 0.52	& 0.05	\\
            &	$-$145--$-135$ 	& 1.05	& 0.08	\\
            &	$-$135--$-125$ 	& 0.85	& 0.06	\\
            &	$-$65--$-55$ 	& 0.35	& 0.05	\\
            &	$-$45--$-35$ 	& 0.53	& 0.04	\\                        
            &	$-$35--$-25$ 	& 0.38	& 0.02	\\
\hline
            \multirow{2}{*}{b} 
            &	$-$75--$-65$	& 0.72	& 0.08	\\
            &	$-$65--$-55$ 	& 1.03	& 0.09	\\            	    	    
\hline
            \multirow{6}{*}{c1} 
            &	$-$90--$-80$	& 0.97	& 0.11	\\
            &	$-$80--$-70$ 	& 1.20	& 0.10	\\
            &	$-$70--$-60$	& 0.92	& 0.10	\\
            &	$-$60--$-50$	& 0.60	& 0.06	\\
	    &	$-$40--$-30$ 	& 0.33	& 0.06	\\
	    &	$-$30--$-20$ 	& 0.37	& 0.09	\\	
\hline
            \multirow{3}{*}{c2} 
            &	$-$100--$-90$	& 1.10	& 0.37	\\
            &	$-$90--$-80$ 	& 0.80	& 0.19	\\
            &	$-$80--$-70$ 	& 0.56	& 0.14	\\
\hline
            \multirow{2}{*}{d} 
            &	$-$70--$-60$	& 1.18	& 0.05	\\
            &	$-$60--$-50$ 	& 1.50	& 0.07	\\
\hline
	    \multirow{9}{*}{e} 
            &	$-$100--$-90$	& 0.92	& 0.03	\\
            &	$-$90--$-80$	& 0.95	& 0.04	\\
            &	$-$80--$-70$ 	& 0.80	& 0.03	\\
            &	$-$70--$-60$	& 0.63	& 0.04	\\
            &	$-$60--$-50$ 	& 0.53	& 0.03	\\
            &	$-$50--$-40$ 	& 0.38	& 0.06	\\
	    &	$-$40--$-30$ 	& 0.33	& 0.02	\\
	    &	0--$10$ 		& 0.35	& 0.05	\\
	    &	10--$20$ 		& 0.62	& 0.05	\\
\hline
\multicolumn{4}{@{}l@{}}{\hbox to 0pt{\parbox{160mm}{
       \footnotemark[$*$] The averaged values in a 10 km s$^{-1}$ winows.
       \par\noindent
   }\hss}}
    \end{tabular}
  \end{center}
\end{table}

\begin{table}[p]
  \caption{LVG results for X(CO)/($dv/dr$) = 1.1 $\times 10^{-5}$ pc (km s$^{-1}$)$^{-1}$}\label{tab:LVGtable}
  \begin{center}
    \begin{tabular}{cccccccccc}
\hline
\hline
\multirow{2}{*}{Peak}  & V$_{\rm LSR}$
           & \multicolumn{2}{c}{$\rm n(H_{2})$ (cm$^{-3}$)}
                      && \multicolumn{2}{c}{$T_{\rm k}$ (K)}  & \multirow{2}{*}{ min. $\chi^2$ } \\

                                 \cline{3-4} \cline{6-7}
           & (km s$^{-1}$)& $\chi^2 < 6.25$ & min. $\chi^2$ 
           && $\chi^2 < 6.25$ & min. $\chi^2$ \\
\hline 
            \multirow{6}{*}{a} 
            &	$-$155--$-145$	& $10^{2.7}$--$10^{3.1}$	& $10^{2.9}$	&& 19--95 		& 42 & $0.4\times 10^{-2}$\\
            &	$-$145--$-135$ 	& $10^{3.6}$ $<$		& $10^{4.4}$	&& 101 $<$ 	& 195 & $1.7\times 10^{-2}$\\
            &	$-$135--$-125$ 	& $10^{3.0}$--$10^{3.7}$	& $10^{3.2}$	&& 47--249 	& 114 & $0.2\times 10^{-2}$\\
            &	$-$65--$-55$ 	& $10^{2.8}$--$10^{3.0}$	& $10^{2.9}$	&&  12--34		& 18 & $0.6\times 10^{-2}$\\
            &	$-$45--$-35$ 	& $10^{2.7}$--$10^{3.0}$	& $10^{2.8}$	&& 34--127 	& 60 & $0.07\times 10^{-2}$\\
            &	$-$35--$-25$ 	& $<$ $10^{2.3}$		& ---			&& 123 $<$ 	& --- & 2.1\\
\hline
            \multirow{2}{*}{b} 
            &	$-$75--$-65$	& $10^{3.1}$--$10^{3.5}$	& $10^{3.2}$	&&  25--86 	& 44 & $0.4\times 10^{-2}$\\
            &	$-$65--$-55$ 	& $10^{3.5}$ $<$		& $10^{4.7}$	&& 70 $<$ 	& 214 & $1\times 10^{-5}$\\
\hline
            \multirow{9}{*}{c1} 
             &	$-$100--$-90$	& ---					& ---			&& --- 		& --- & 31.2\\
            &	$-$90--$-80$ 	& ---					& ---			&& --- 		& --- & 5.8\\
            &	$-$80--$-70$ 	& $10^{2.8}$--$10^{3.1}$	& $10^{2.9}$	&& 167 $<$ 	& --- & $3.4\times 10^{-2}$\\
            &	$-$70--$-60$	& $10^{2.6}$--$10^{2.9}$	& $10^{2.7}$	&& 87 $<$ 	& 194 & $0.3\times 10^{-2}$\\
            &	$-$60--$-50$ 	& $10^{2.4}$--$10^{2.8}$	& $10^{2.6}$	&& 53 $<$ 	& 128 & $0.01\times 10^{-2}$\\
            &	$-$50--$-40$ 	& $10^{2.0}$--$10^{2.7}$	& $10^{2.2}$	&& 37 $<$	 	& 188 & $1 \times 10^{-5}$\\
	    &	$-$40--$-30$ 	& $<$ $10^{2.6}$		& ---			&&  31 $<$  	& 245 & 1.3\\
	    &	0--$10$ 		& $10^{2.8}$--$10^{3.0}$	& $10^{2.9}$	&& 12--26	 	& 16 & $1.7\times 10^{-2}$\\\
	    &	10--$20$ 		& $10^{2.6}$--$10^{3.6}$	& $10^{3.0}$	&& 32--255 	& 48 & $0.05\times 10^{-2}$\\
\hline
            \multirow{3}{*}{c2} 
            &	$-$100--$-90$	& ---					& ---			&& --- 		& --- & 60.0\\
            &	$-$90--$-80$ 	& ---					& ---			&& --- 		& --- & 7.0\\
            &	$-$80--$-70$ 	& $10^{3.4}$ $<$		& $10^{3.8}$	&&  21--190	& 66 & $0.2\times 10^{-2}$\\
\hline
            \multirow{2}{*}{d} 
            &	$-$70--$-60$	& $10^{3.4}$--$10^{4.5}$	& ---			&& 245 $<$ 	 & --- & 0.4\\
            &	$-$60--$-50$ 	& ---					& ---			&& --- 		& --- & 14.5\\
\hline            
	    \multirow{6}{*}{e} 
            &	$-$90--$-80$	& $10^{3.5}$ $<$		& $10^{3.9}$	&& 45--280 	& 94 & $0.08\times 10^{-2}$\\
            &	$-$80--$-70$ 	& ---					& ---			&& --- 		& --- & 5.8\\
            &	$-$70--$-60$ 	& $10^{3.4}$ $<$		& $10^{3.7}$	&& 45--286 	& 82 & $0.04\times 10^{-2}$\\
            &	$-$60--$-50$	& $10^{3.0}$ -- $10^{3.2}$	& $10^{3.1}$	&& 22--71 		& 36 & $0.03\times 10^{-2}$\\
            &	$-$40--$-30$ 	& $10^{2.8}$ -- $10^{3.1}$	& $10^{3.0}$	&& $<$ 24 	& 14 & $1.8\times 10^{-2}$ \\
	    &	$-$30--$-20$ 	& $10^{3.1}$ -- $10^{3.3}$ & $10^{3.2}$ 	&& $<$ 15		& --- & $0.3\times 10^{-2}$\\
\hline

    \end{tabular}
  \end{center}
\end{table}

\begin{table}[p]
  \caption{Physical properties of the compact broad components in panel b of Figures \ref{fig:channelCO32loop1eastfp}, \ref{fig:channelCO32loop1westfp}, \ref{fig:channelCO32loop2westfp} and \ref{fig:channelCO32loop3westfp}}\label{tab:cloudtable}
  \begin{center}
    \begin{tabular}{cccccccccc}
\hline
\hline

Region& no.  & $W_{\mathrm{CO}(J=3-2)}$ ($10^3$ K km s$^{-1}$)
           && $R_{3-2/2-1}$  &&$R_{3-2/1-0}$ \footnotemark[$\dagger$]  &Mass ($10^3 \MO$) \footnotemark[$\dagger$]\\
\hline

	    \multirow{2}{*}{A} 
            &1 &	 0.5	&& 0.23 &&  --- & --- \\
            &2 &	 6.6 	&& 0.40 &&  --- & --- \\

\hline
	    \multirow{4}{*}{C} 
            &3 &	 3.5 	&& 0.77 &&  --- &  --- \\
            &4 &	 0.4 	&&  ---\footnotemark[$*$] &&  0.58 & 2.0 \\
            &5 &	 0.6	&& 0.93 &&  0.46 & 1.5 \\
            &6 &	 2.2 	&& 0.90 &&  --- &  --- \\
\hline
	    \multirow{3}{*}{E} 
            &7 &	 5.8	&& --- &&  --- &  --- \\
            &8 &	 2.1 	&& 0.80 &&  --- & --- \\
            &9 &	 4.1	&& 0.73 &&  --- & --- \\
\hline
	    \multirow{5}{*}{G} 
            &10 &	 4.8	&& --- &&  --- &  --- \\
            &11 &	 4.0	&&  0.50 &&  --- & --- \\
            &12 &	 0.7	&& 0.55 &&  --- & --- \\
            &13 &	 0.7	&& 0.52 &&  --- & ---\\
            &14 &	 1.4 	&& 0.55 &&  --- &  ---\\
\hline
 \multicolumn{4}{@{}l@{}}{\hbox to 0pt{\parbox{160mm}{\footnotesize
Notes. '-' means no value because of outside of the observing region in the line. 
       \par\noindent
       \footnotemark[$*$] $^{12}$CO($J=$2--1) is not detected at more than 3$\sigma$ level.
      \par\noindent
      \footnotemark[$\dagger$]  Using Mopra $^{12}$CO($J=$1--0) data (T10b). 
       \par\noindent
   }\hss}}
    \end{tabular}
  \end{center}
\end{table}

\newpage

\clearpage

\clearpage

\begin{figure}
  \begin{center}
    \FigureFile(100mm,100mm){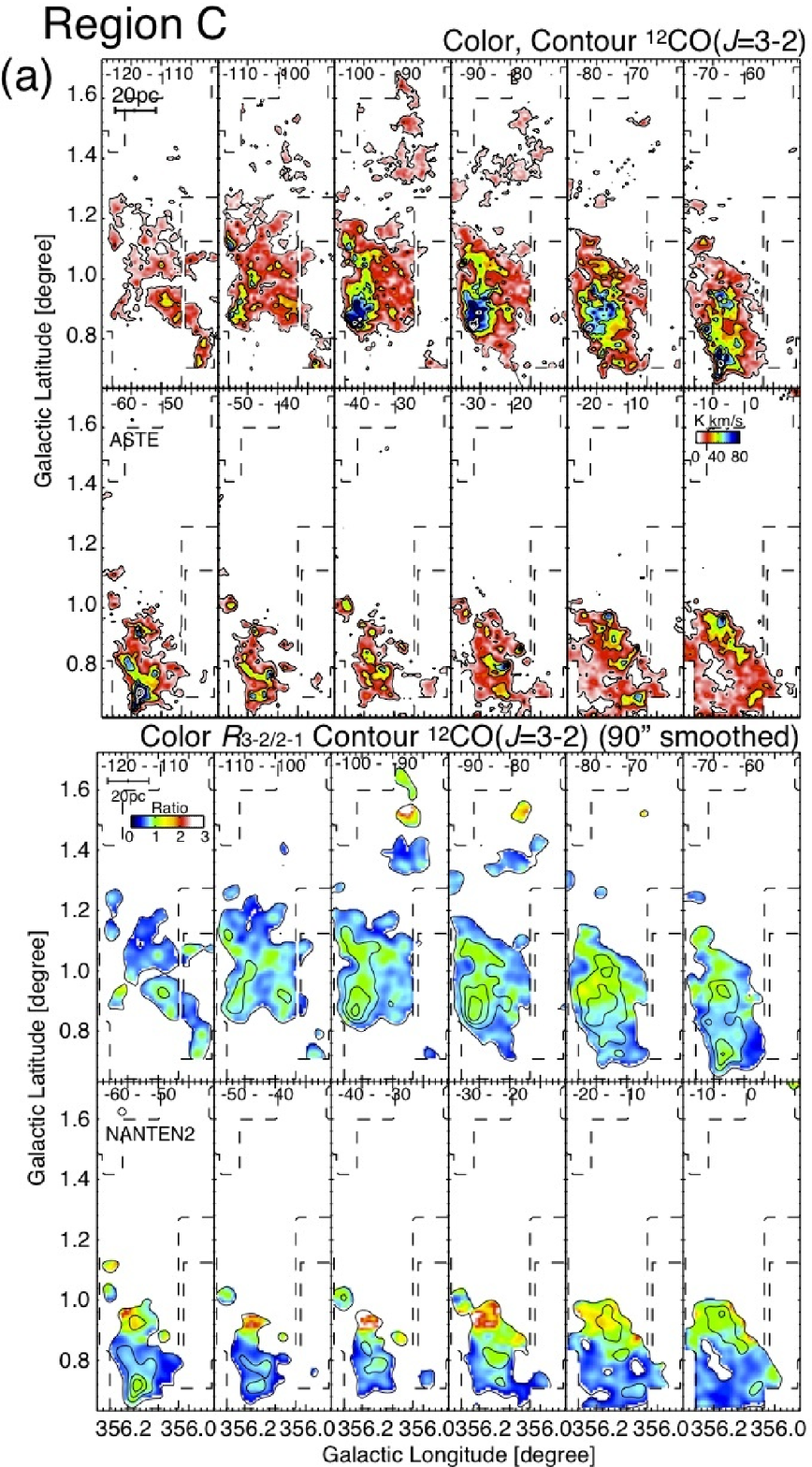}
  \end{center}
  \caption{(a-top)Velocity channel distributions of $^{12}$CO($J$=3--2) integrated intensity every 10 km s$^{-1}$ at the west footpoint of loop 1. Contours are illustrated every 20 K km s$^{-1}$ from 6 K km s$^{-1}$. (a-bottom)Velocity channel distributionss of $R_{3-2/2-1}$ integrated integrated intensity every 10 km s$^{-1}$. Contours show the $^{12}$CO($J$=3--2) emission smoothed with a gaussian function to a 90$''$ spatial resolution and are plotted every 20 K km s$^{-1}$ from 6 K km s$^{-1}$.}
  \label{fig:channelCO32loop1westfp}
\end{figure}

\begin{figure}
  \begin{center}
    \FigureFile(160mm,160mm){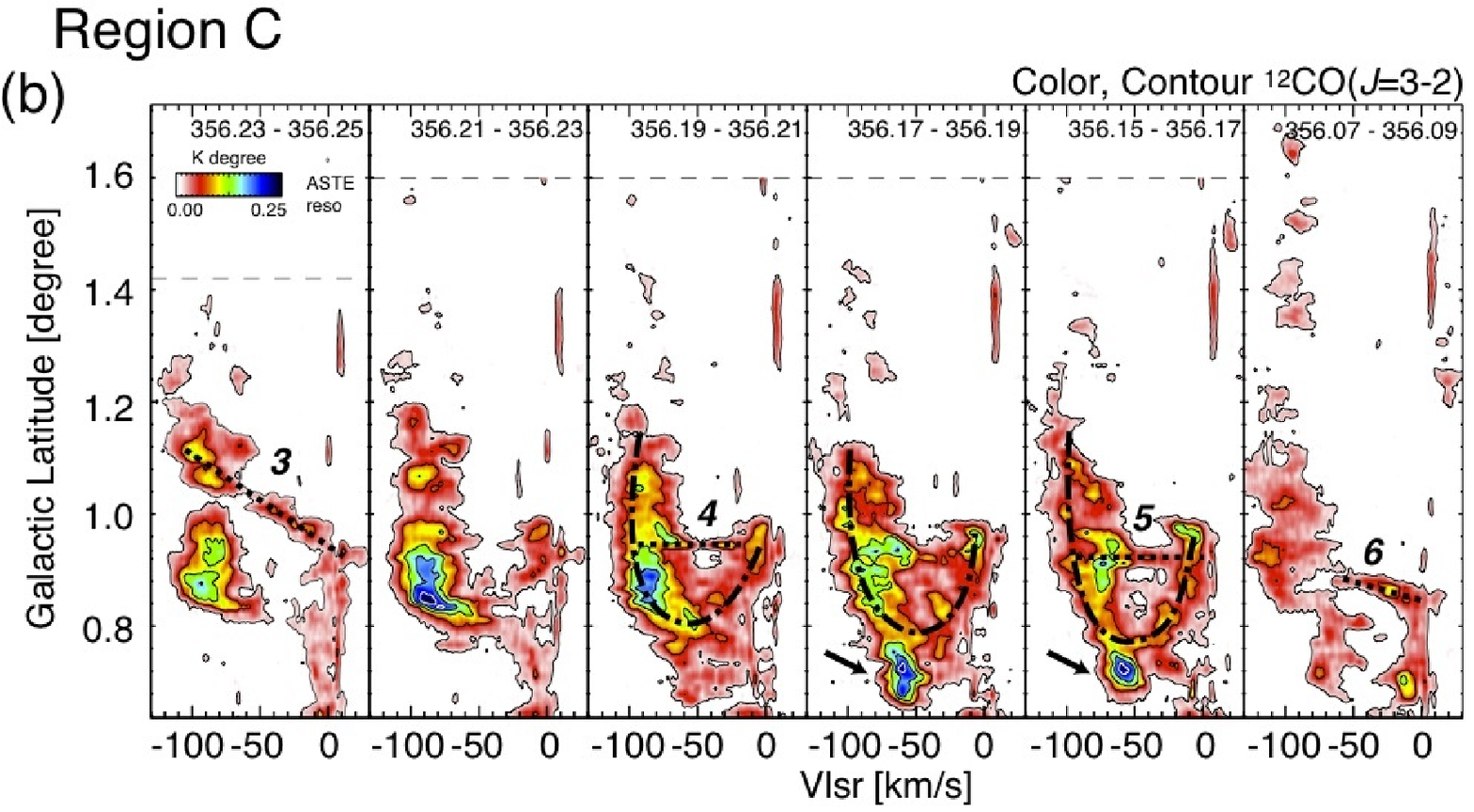}
  \end{center}
  \contcaption{(b)Longitude velocity distributions of the $^{12}$CO($J$=3--2) with interval of 100$''$ in $l$ in regions C. Contours are plotted every 0.05 K deg. from 0.02 K deg..}
  \label{fig:channelCO32loop1westfp_2}
\end{figure}

\begin{figure}
  \begin{center}
    \FigureFile(80mm,100mm){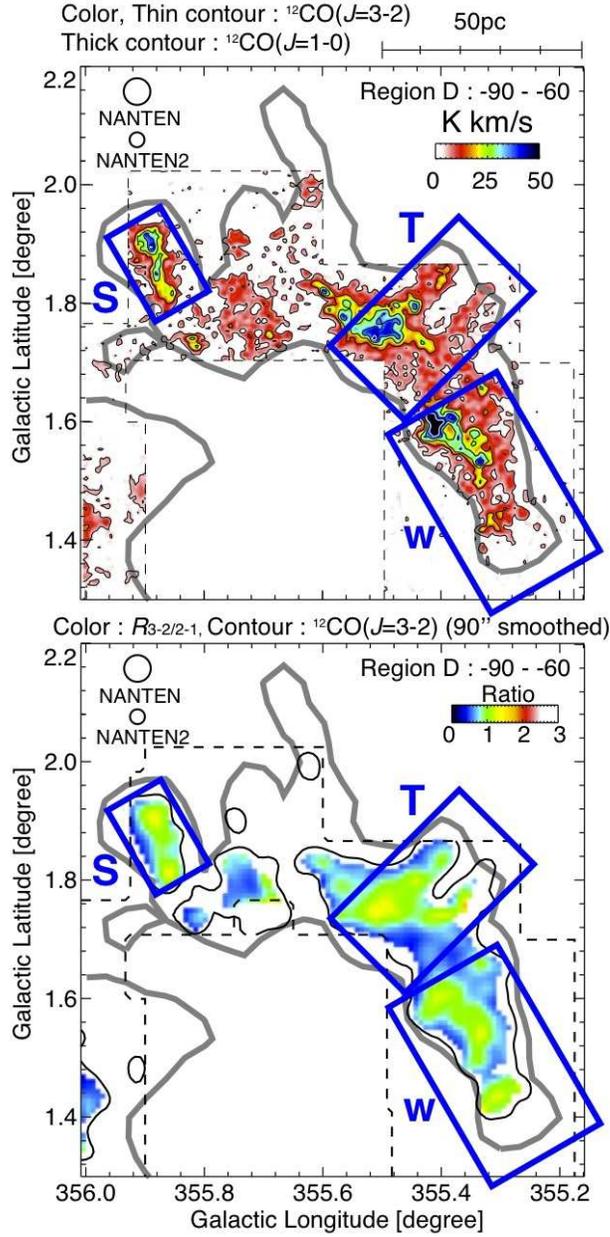}
  \end{center}
  \caption{(top) Integrated Intensity distributions of loop 2 in $^{12}$CO($J$=3--2), color and thin contours, and $^{12}$CO($J$=1--0), thick contours, in region D. The integration range in velocity is from $-90$ km s$^{-1}$ to $-60$ km s$^{-1}$. Dashed lines indicate the observed regions. (bottom) Distributions of the $R_{3-2/2-1}$. Contours show the $^{12}$CO($J$=3--2) emission smoothed with a gaussian function to a 90$''$ spatial resolution and are plotted every 10 K km s$^{-1}$ 4 K km s$^{-1}$.}
  \label{fig:channelCO32loop2top}
\end{figure}

\begin{figure}
  \begin{center}
    \FigureFile(130mm,130mm){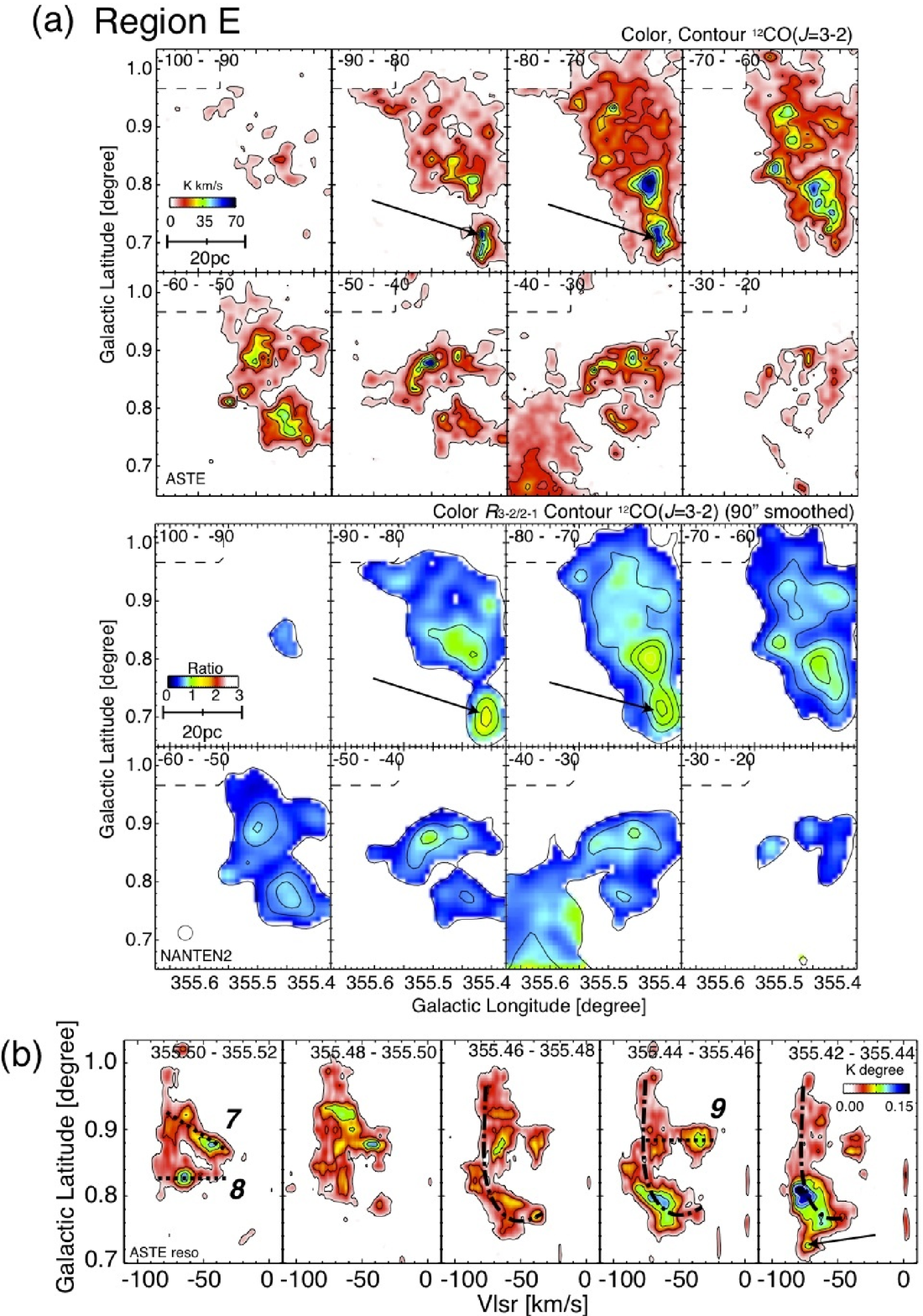}
  \end{center}
  \caption{(a-top) Velocity channel distributions of $^{12}$CO($J$=3--2) integrated intensity every 10 km s$^{-1}$ at the west footpoint of loop 2. Contours are illustrated every 10 K km s$^{-1}$ from 4 K km s$^{-1}$. (a-bottom)Velocity channel distributions of the $R_{3-2/2-1}$ integrated every 10 km s$^{-1}$. Contours show the $^{12}$CO($J$=3--2) emission smoothed with a gaussian function to a 90$''$ spatial resolution and are plotted every 10 K km s$^{-1}$ 4 K km s$^{-1}$. (b)Longitude velocity distributions of the $^{12}$CO($J$=3--2) with interval of 100$''$ in $l$ in regions E. Contours are plotted every 0.05 K deg. from 0.02 K deg..}
  \label{fig:channelCO32loop2westfp}
\end{figure}

\begin{figure}
  \begin{center}
    \FigureFile(160mm,160mm){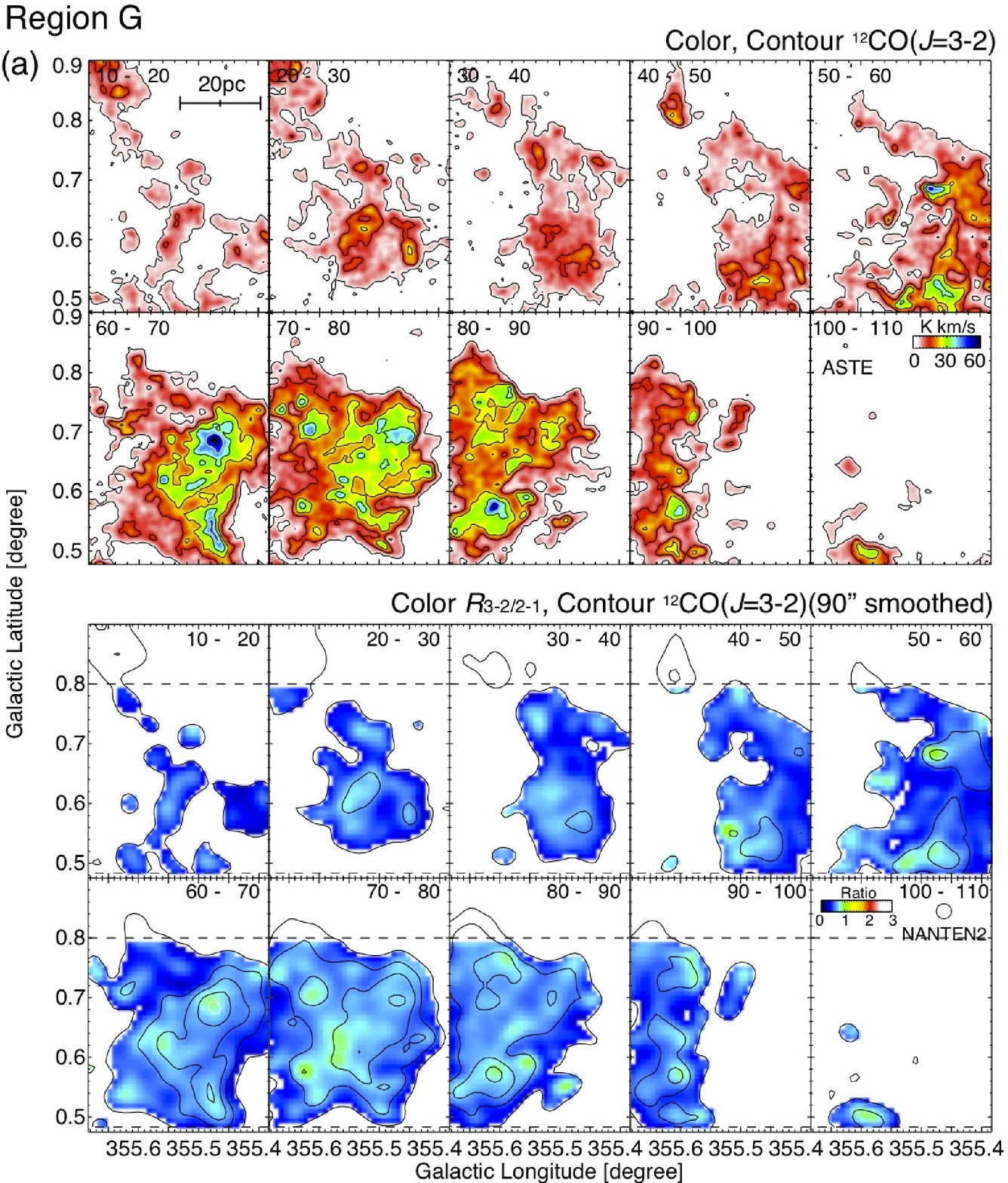}
  \end{center}
  \caption{(top)Velocity channel distributions of $^{12}$CO($J$=3--2) integrated intensity every 10 km s$^{-1}$ at the west footpoint of loop 3. Contours are illustrated every 10 K km s$^{-1}$ from 4 K km s$^{-1}$. (bottom)Velocity channel distributions of the $R_{3-2/2-1}$ intensity ratio integrated every 10 km s$^{-1}$. Contours show the $^{12}$CO($J$=3--2) emission that was smoothed with a Gaussian function to a 90¢òh spatial resolution and are plotted every 10 K km s$^{-1}$ from 4 K km s$^{-1}$.}
  \label{fig:channelCO32loop3westfp}
\end{figure}

\begin{figure}
  \begin{center}
    \FigureFile(160mm,160mm){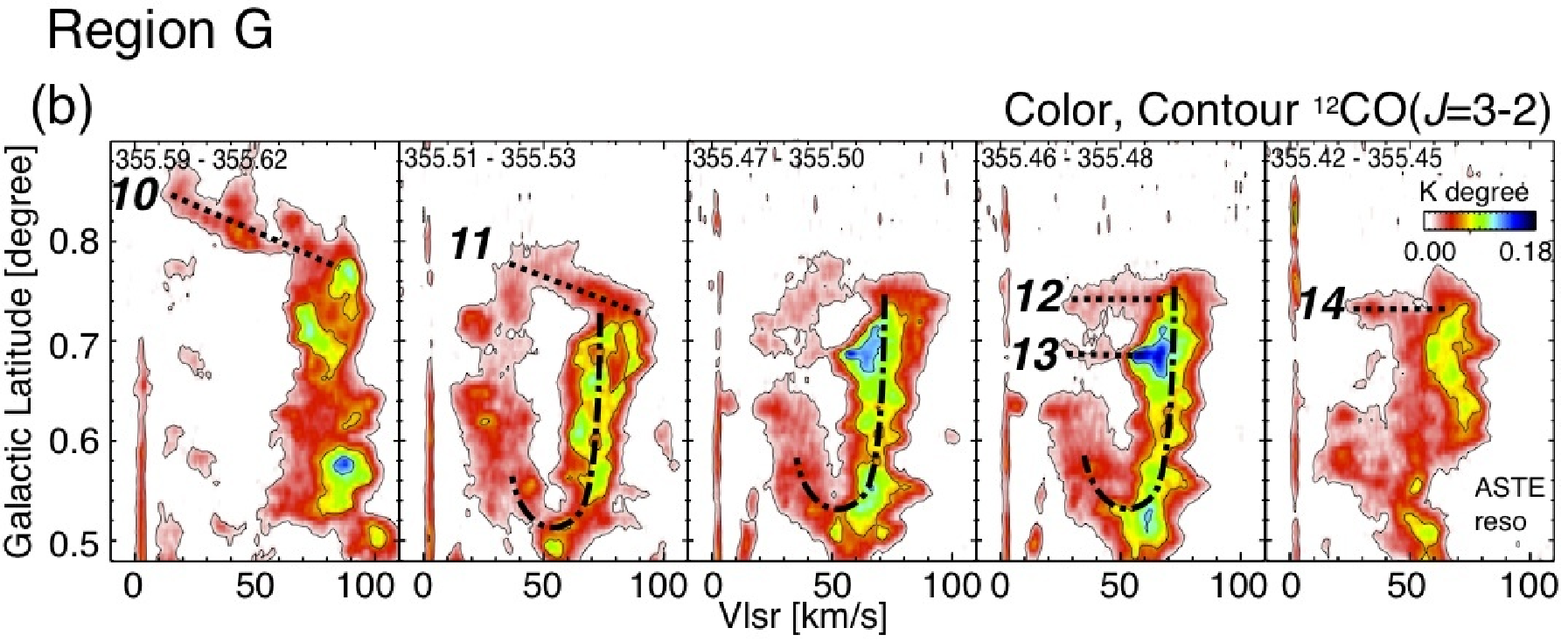}
  \end{center}
  \contcaption{(b)Longitude velocity distributions of the $^{12}$CO($J$=3--2) with interval of 100$''$ in $l$ in regions G. Contours are plotted every 0.05 K deg. from 0.02 K deg..}
  \label{fig:channelCO32loop1westfp_2}
\end{figure}

\clearpage

\begin{figure}[p]
  \begin{center}
    \FigureFile(120mm,120mm){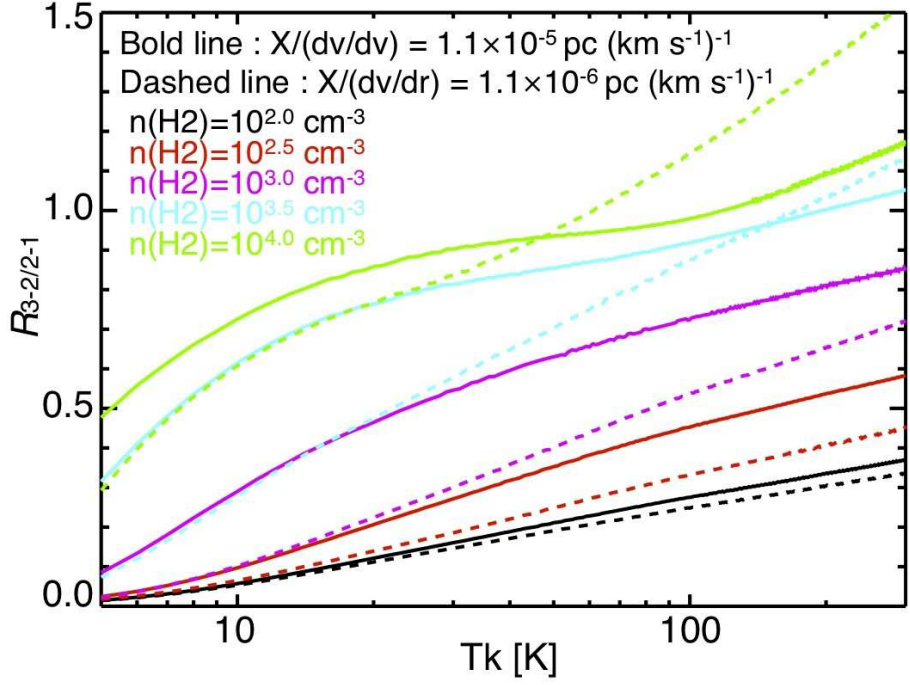}
  \end{center}
  \caption{$R_{3-2/2-1}$ as a function of kinetic temperature $T_\mathrm{k}$ estimated by the LVG calculations. We take two X(CO)/($dv$/$dr$), $1.1 \times 10^{-5}$(Bold line) and $1.1 \times 10^{-6}$(Dashed line). Colored lines show densities of 10$^{2.0}$(green), 10$^{2.5}$(blue), 10$^{3.0}$(purple), 10$^{3.5}$(red) and 10$^{4.0}$ cm$^{-3}$(black).}
  \label{fig:LVGplotconstdensity}
\end{figure}

\clearpage
\begin{figure}[p]
  \begin{center}
    \FigureFile(130mm,130mm){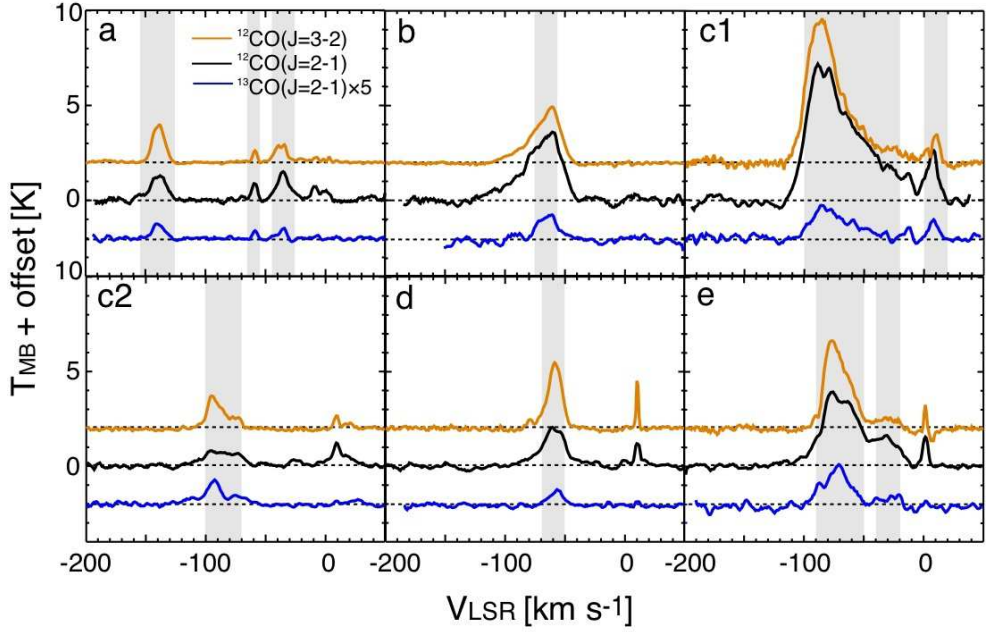}
  \end{center}
  \caption{CO spectra at 6 positions shown in Figure \ref{fig:ratiolb}. The intensities of the$^{13}$CO($J$=2--1) emission were multiplied by 5.}
  \label{fig:LVGspectra}
\end{figure}

\begin{figure}[p]
  \begin{center}
    \FigureFile(130mm,130mm){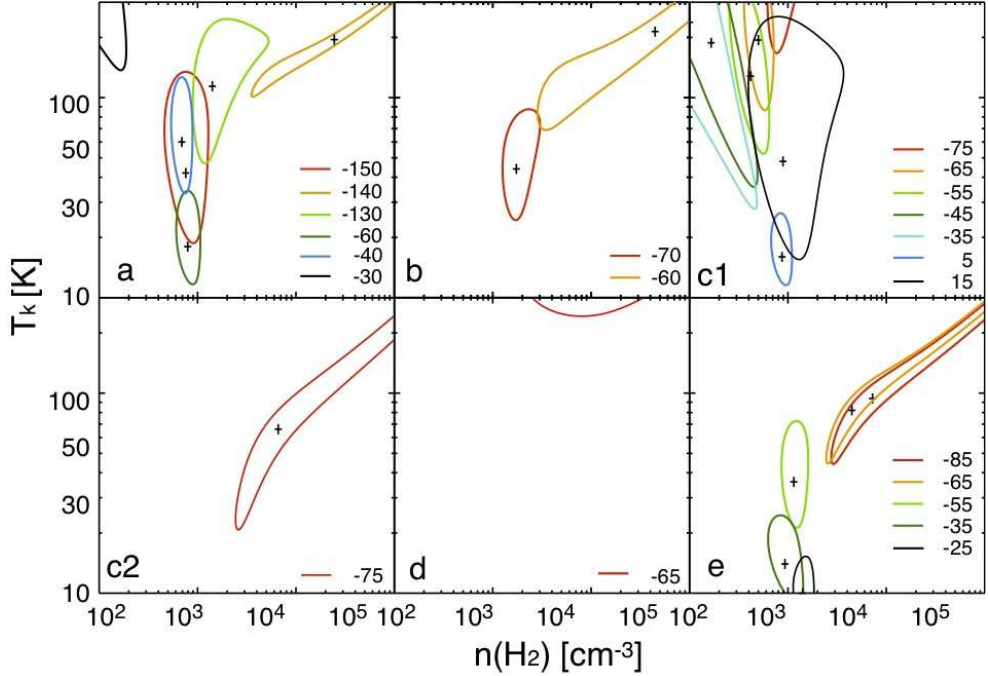}
  \end{center}
  \caption{Chi square, $\chi^{2}$, distributions estimated from the LVG calculations on the temperature-density plane for the different velocity of the spectra shown in Figure \ref{fig:LVGspectra}. Colored lines are plotted at $\chi^{2}$ of 3.84, which corresponds to a 95\% confidence level, and the crosses show the points that minimum $\chi^{2}$ is found.}
  \label{fig:LVGresult}
\end{figure}

\begin{figure}[p]
  \begin{center}
    \FigureFile(160mm,160mm){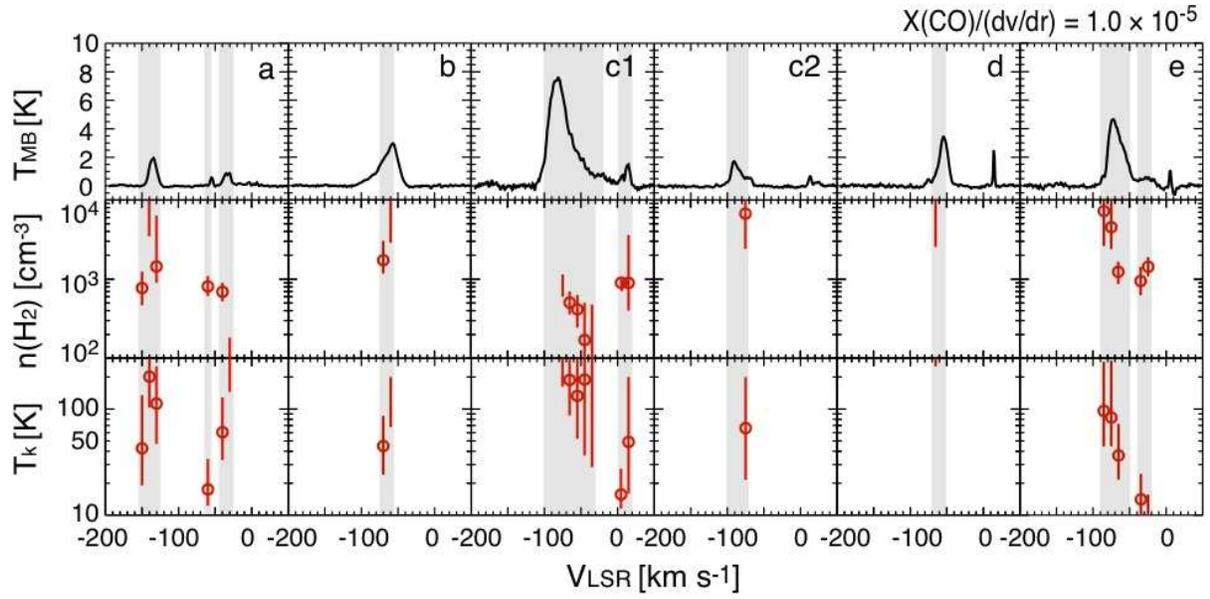}
  \end{center}
  \caption{LVG results for X(CO)/($dv/dr$) of $1.1 \times 10^{-5}$ for six spectra shown in Figure \ref{fig:LVGspectra}. Horizontal axis of all figures is velocity. Top row shows the $^{12}$CO($J$=3--2) spectra at the peaks a-f shown in Figure \ref{fig:ratiolb}. The second and third rows show number density, $n\rm(H_2)$, and kinetic temperature, $T_k$, respectively. Open circles show the lowest point of $\chi^{2}$, and the error range is defined as 5\% confidence level of $\chi^{2}$ distribution with 1 degree of freedom, which corresponds to $\chi^{2} = 3.84$.}
  \label{fig:LVGresultplot}
\end{figure}

\clearpage

\begin{figure}
  \begin{center}
    \FigureFile(160mm,160mm){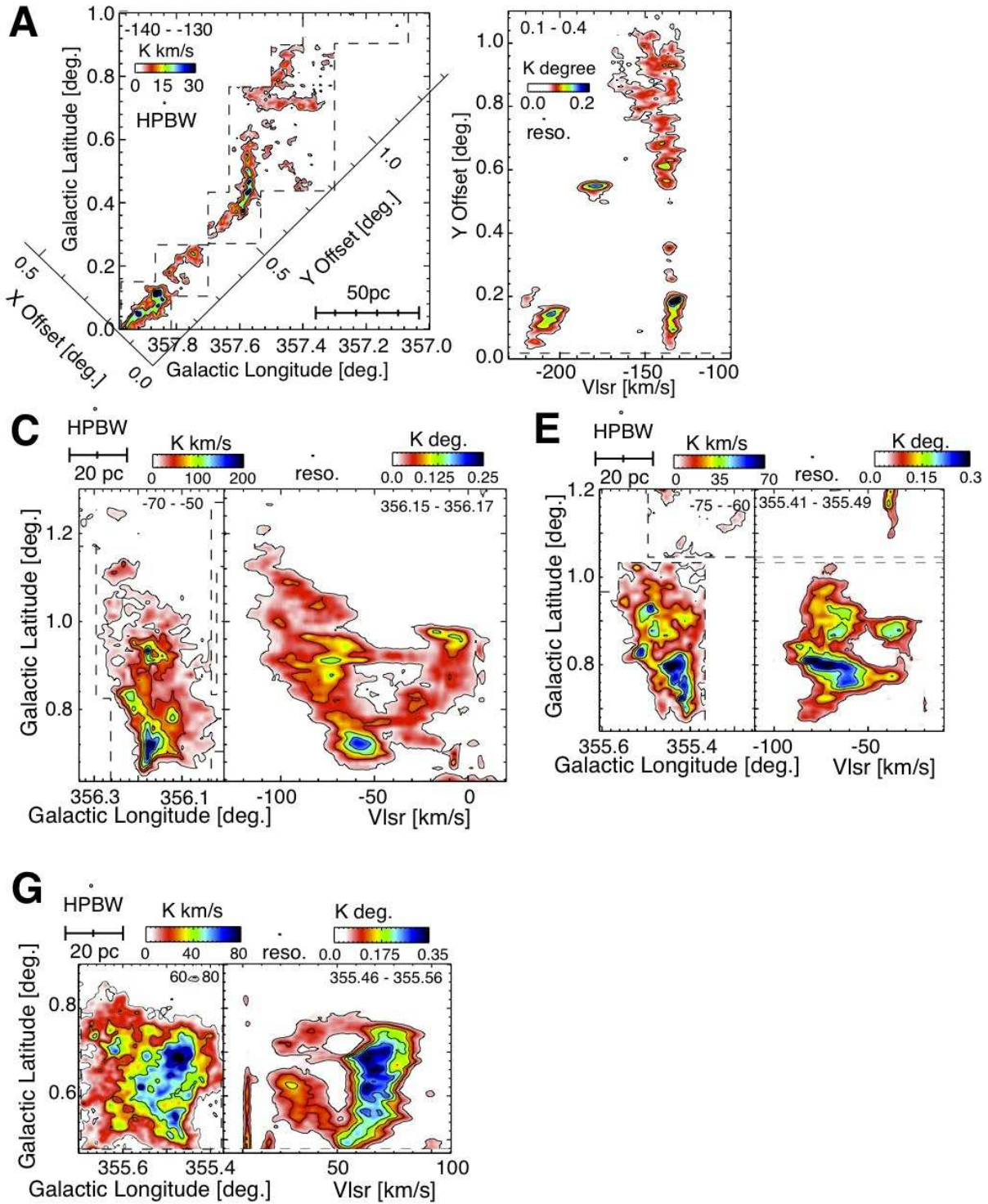}
  \end{center}
  \caption{The sets of the longitude-latitude distributions (left panels) and velocity-latitude distributions (right panels) in $^{12}$CO($J=$3--2) at the four footpoints of the loops. In the Region A, the right panel is the velocity distributions for the X-axis which was defined in the left panel of the figure.}
  \label{fig:fplistCO32}
\end{figure}

\begin{figure}
  \begin{center}
    \FigureFile(135mm,135mm){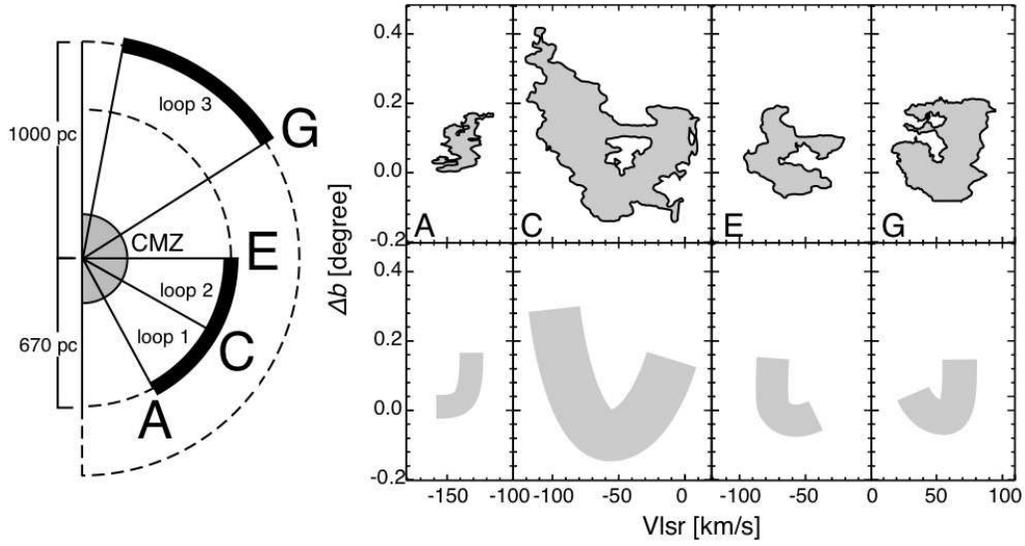} 
  \end{center}
  \caption{(left) A schematic image of the positions of loops 1, 2 and 3 in a face on view. (right) Latitude - velocity diagram of the four foot
points. top-panels: Outer boundaries of the foot points are shown. Origins of the perpendicular axis of the regions A, C, E and G are
$b=$\timeform{0.D68}, \timeform{0.D8}, \timeform{0D.8} and \timeform{0.D56}, respectively. bottom-panels: Schematic images of U shape. }
  \label{fig:fplist+shc}
\end{figure}

\begin{figure}
\renewcommand{\thefigure}{A}
 \begin{center}
    \FigureFile(160mm,160mm){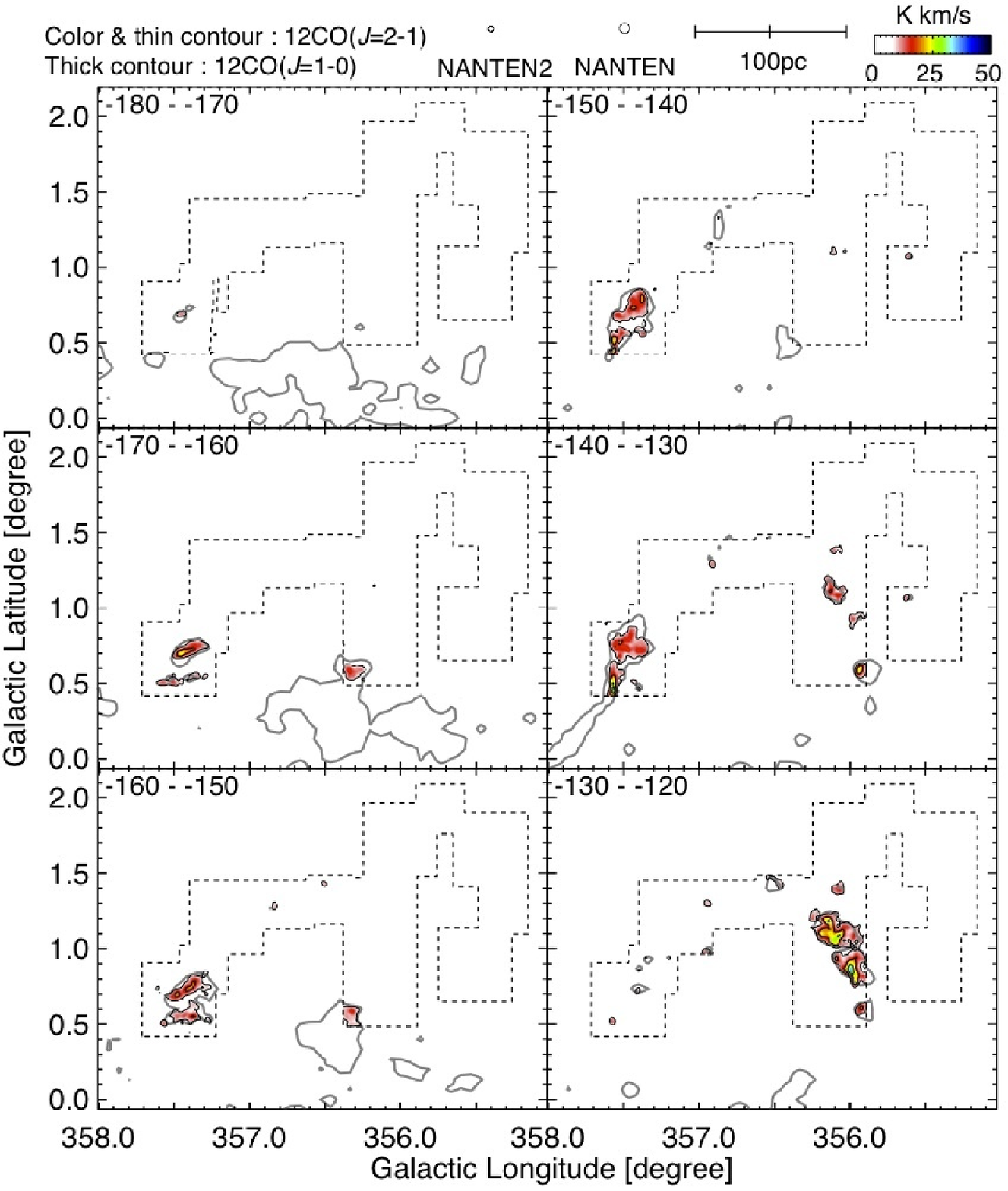}
  \end{center}
  \caption{Velocity-channel distributions of the $^{12}$CO($J=$2--1) emission integrated every 10 km s$^{-1}$ from $-180$ to 60 km s$^{-1}$. Color and thin contours (black and white) show the $^{12}$CO($J=$2--1) emission. The contours are plotted every 6 K km s$^{-1}$(black) and 12 K km s$^{-1}$(white) from 4.7 K km s$^{-1}$. Thick contours show the $^{12}$CO($J=$1--0) emission and are plotted at 7 K km s$^{-1}$. The numbers shown in the left-top of the each panel show the integration range in km s$^{-1}$. Dashed-lines indicate the observed region.}
  \label{fig:channelCO10vsCO21_1}
\end{figure}

\begin{figure}
\renewcommand{\thefigure}{A}
  \begin{center}
    \FigureFile(160mm,160mm){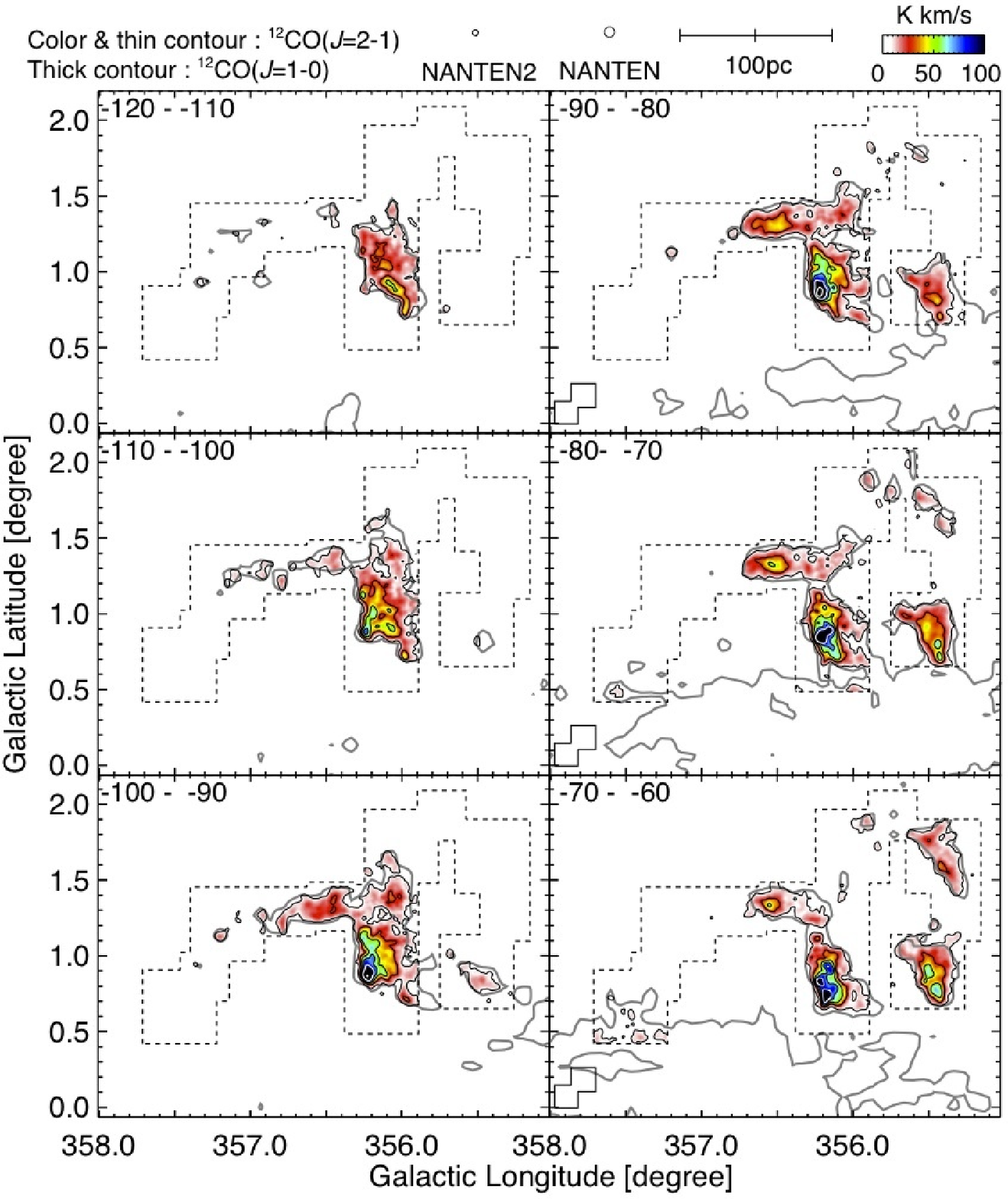}
  \end{center}
  \contcaption{Continued, but the levels of thin contours are different from the previous figure. The contours are plotted every 12 K km s$^{-1}$(black) and 24 K km s$^{-1}$(white) from 4.7 K km s$^{-1}$.}
\end{figure}

\begin{figure}
\renewcommand{\thefigure}{A}
  \begin{center}
    \FigureFile(160mm,160mm){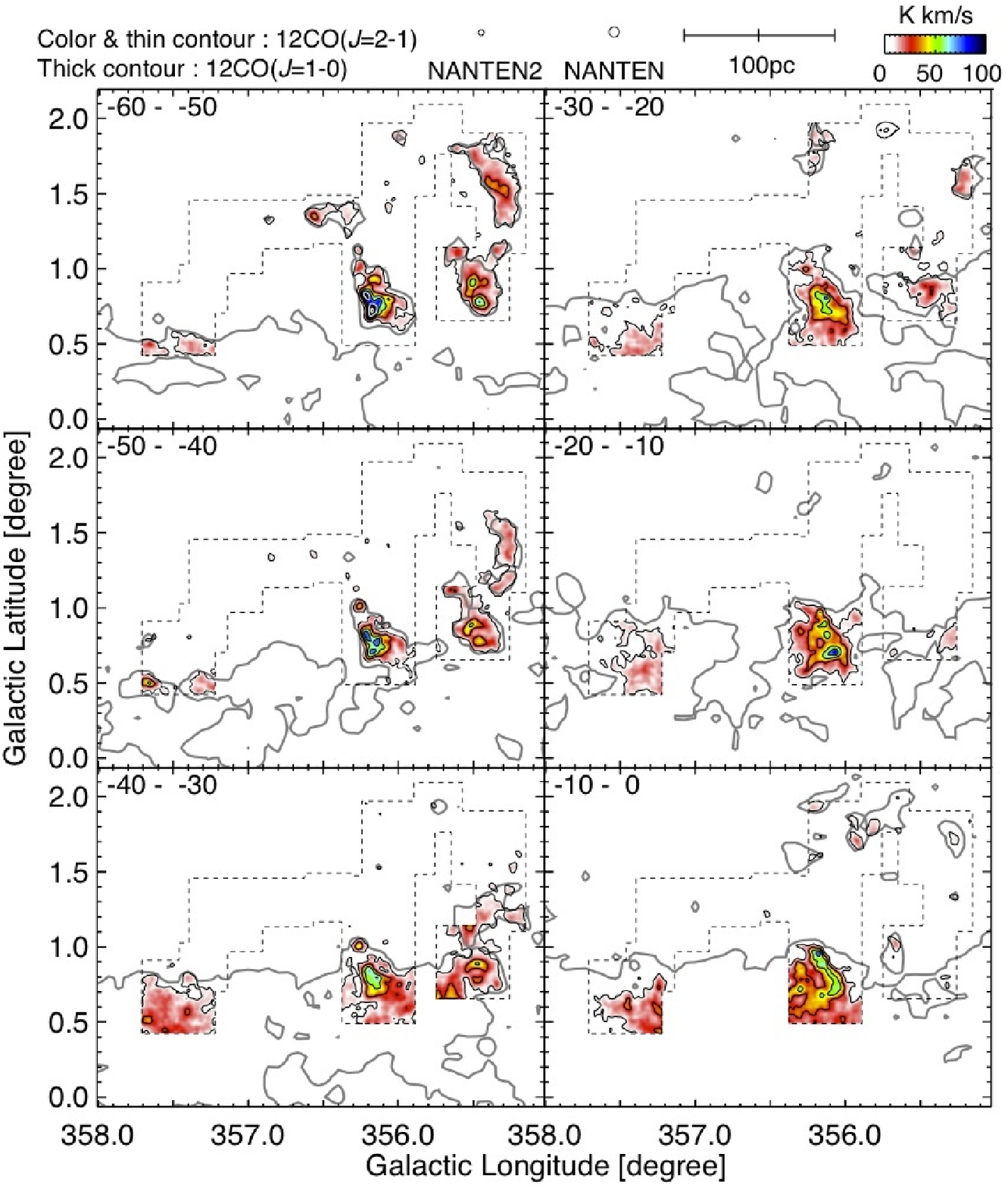}
  \end{center}
  \contcaption{Continued.}
  \label{fig:channelCO10vsCO21_3}
\end{figure}

\begin{figure}
\renewcommand{\thefigure}{A}
  \begin{center}
    \FigureFile(160mm,160mm){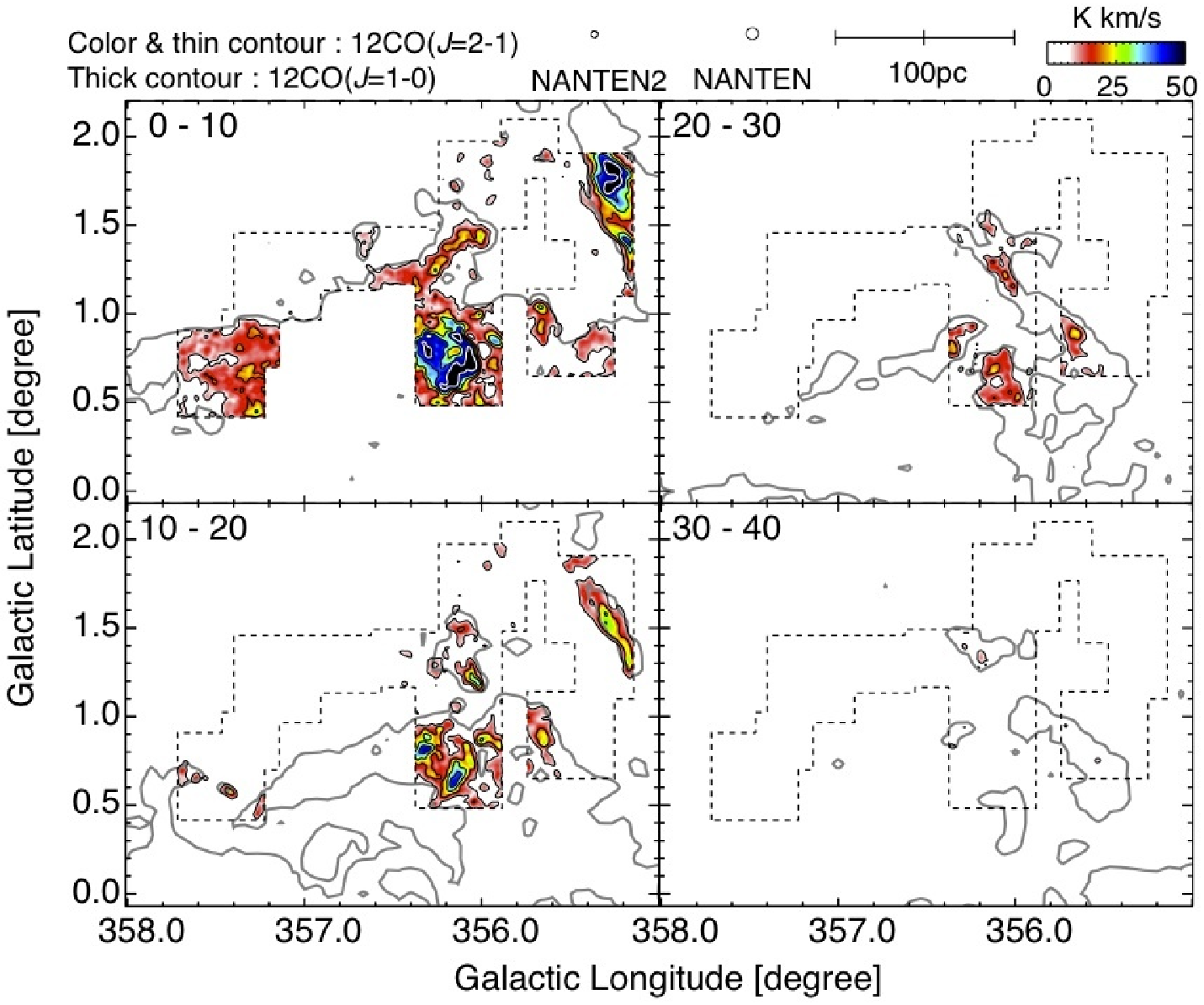}
  \end{center}
  \contcaption{Velocity-channel distributions in $^{12}$CO($J=$2--1). 
Thin black and white contours are $^{12}$CO($J=$2--1), plotted every 24 K km s$^{-1}$(black) and 48 K km s$^{-1}$(white) from 4.7 K km s$^{-1}$. Dashed-lines indicated the observed region.}
  \label{fig:channelCO10vsCO21_4}
\end{figure}

\clearpage

\begin{figure}
\renewcommand{\thefigure}{B}
  \begin{center}
    \FigureFile(160mm,160mm){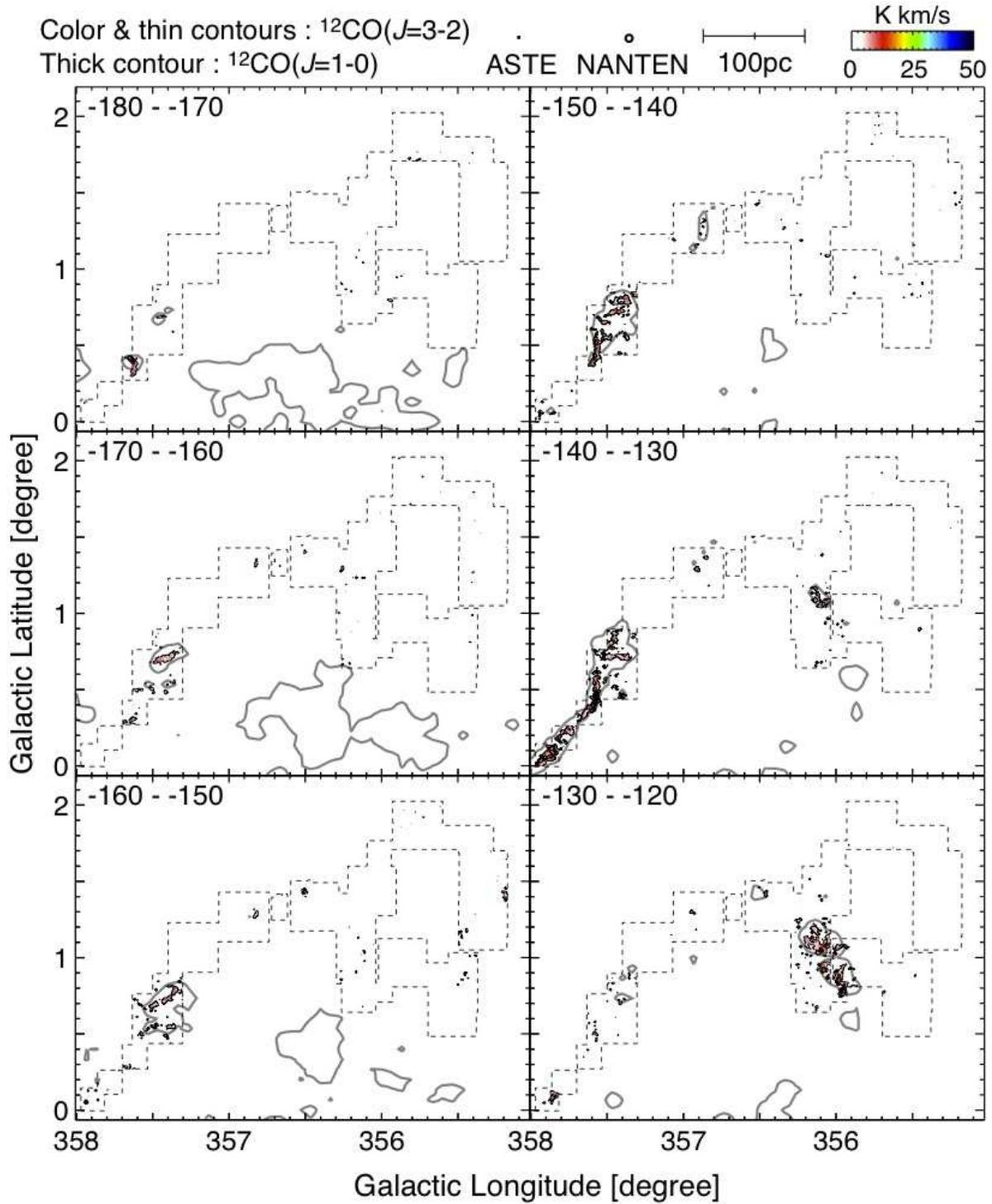}
  \end{center}
  \caption{Velocity-channel distributions of the $^{12}$CO($J=$3--2) emission integrated every 10 km s$^{-1}$ from $-180$ to 120 km s$^{-1}$. Color and thin contours show the $^{12}$CO($J=$3--2) emission, and the contours are plotted every 10 K km s$^{-1}$ from 4 K km s$^{-1}$. Thick contours show the $^{12}$CO($J=$1--0) emission and are plotted 7 K km s$^{-1}$. The numbers shown in the left-top of the each panel show the integration range in km s$^{-1}$. Dashed-lines indicate the observed regions.}\label{fig:channelCO10vsCO32_1}
\end{figure}

\begin{figure}
\renewcommand{\thefigure}{B}
  \begin{center}
    \FigureFile(160mm,160mm){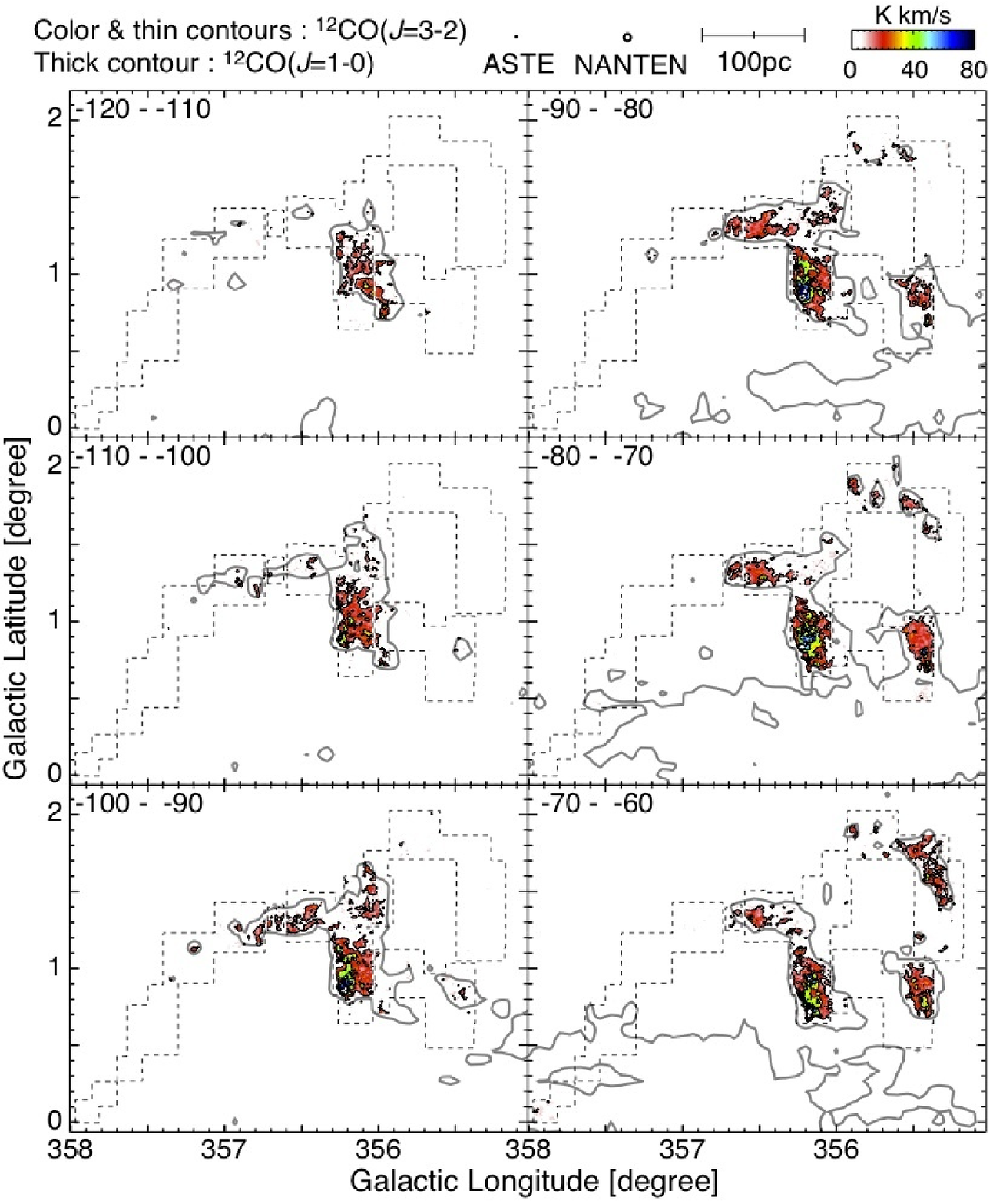}
  \end{center}
  \contcaption{Continued, but the levels of the thin contours are different from the previous figure, that are plotted every 20 K km s$^{-1}$ from 4 K km s$^{-1}$.}
  \label{fig:channelCO10vsCO32_2}
\end{figure}

\begin{figure}
\renewcommand{\thefigure}{B}
  \begin{center}
    \FigureFile(160mm,160mm){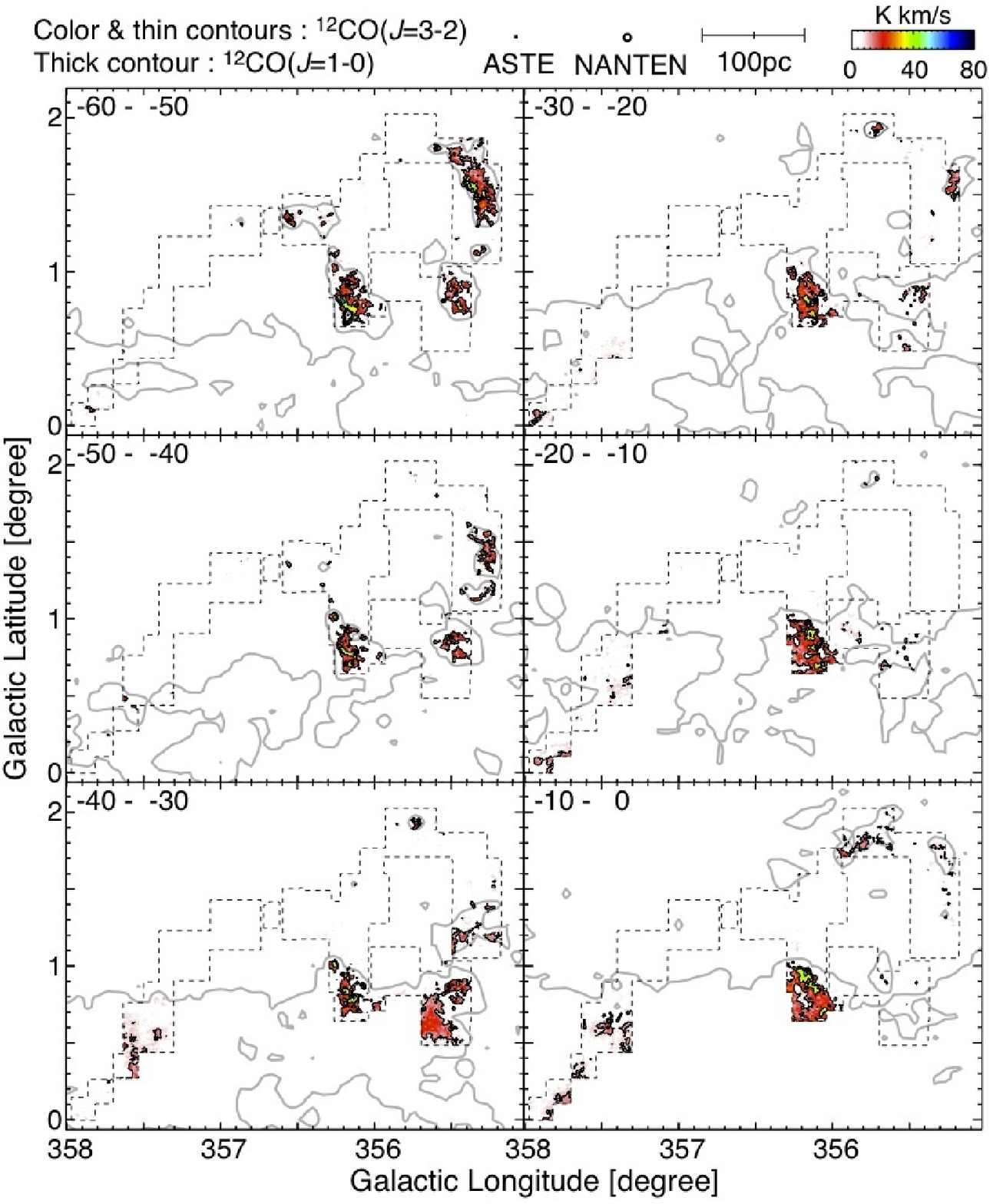}   
  \end{center}
  \contcaption{Continued.}
  \label{fig:channelCO10vsCO32_3}
\end{figure}

\begin{figure}
  \renewcommand{\thefigure}{B}
  \begin{center}
    \FigureFile(160mm,160mm){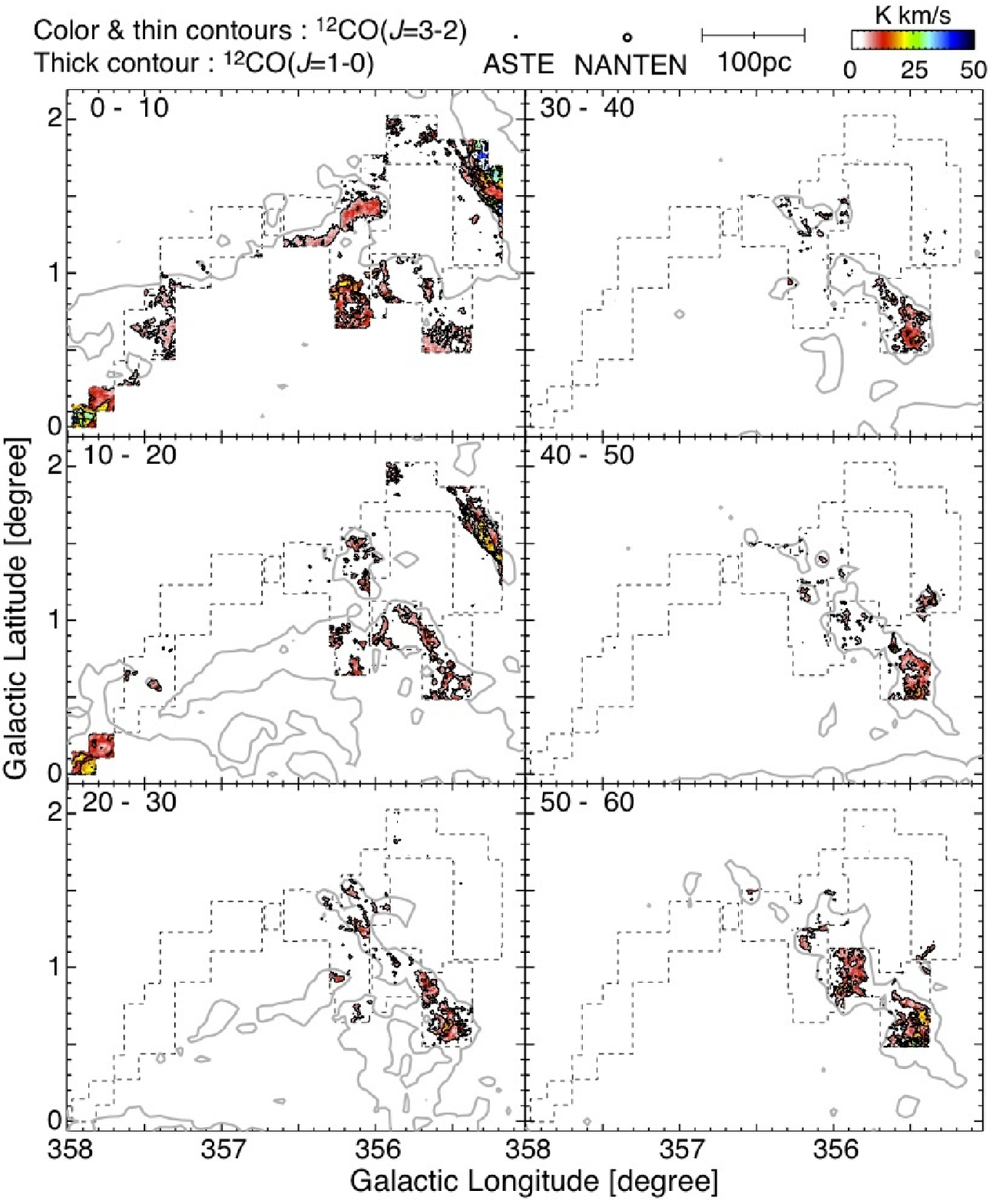}   
  \end{center}
  \contcaption{Color and thin contours are $^{12}$CO($J=$3--2), plotted every 10 K km s$^{-1}$ from 4 K km s$^{-1}$ and thick contours are $^{12}$CO($J=$1--0), plotted 7 K km s$^{-1}$. Dashed-lines indicate the observed regions.}
  \label{fig:channelCO10vsCO32_4}
\end{figure}

\begin{figure}
\renewcommand{\thefigure}{B}
  \begin{center}
    \FigureFile(160mm,160mm){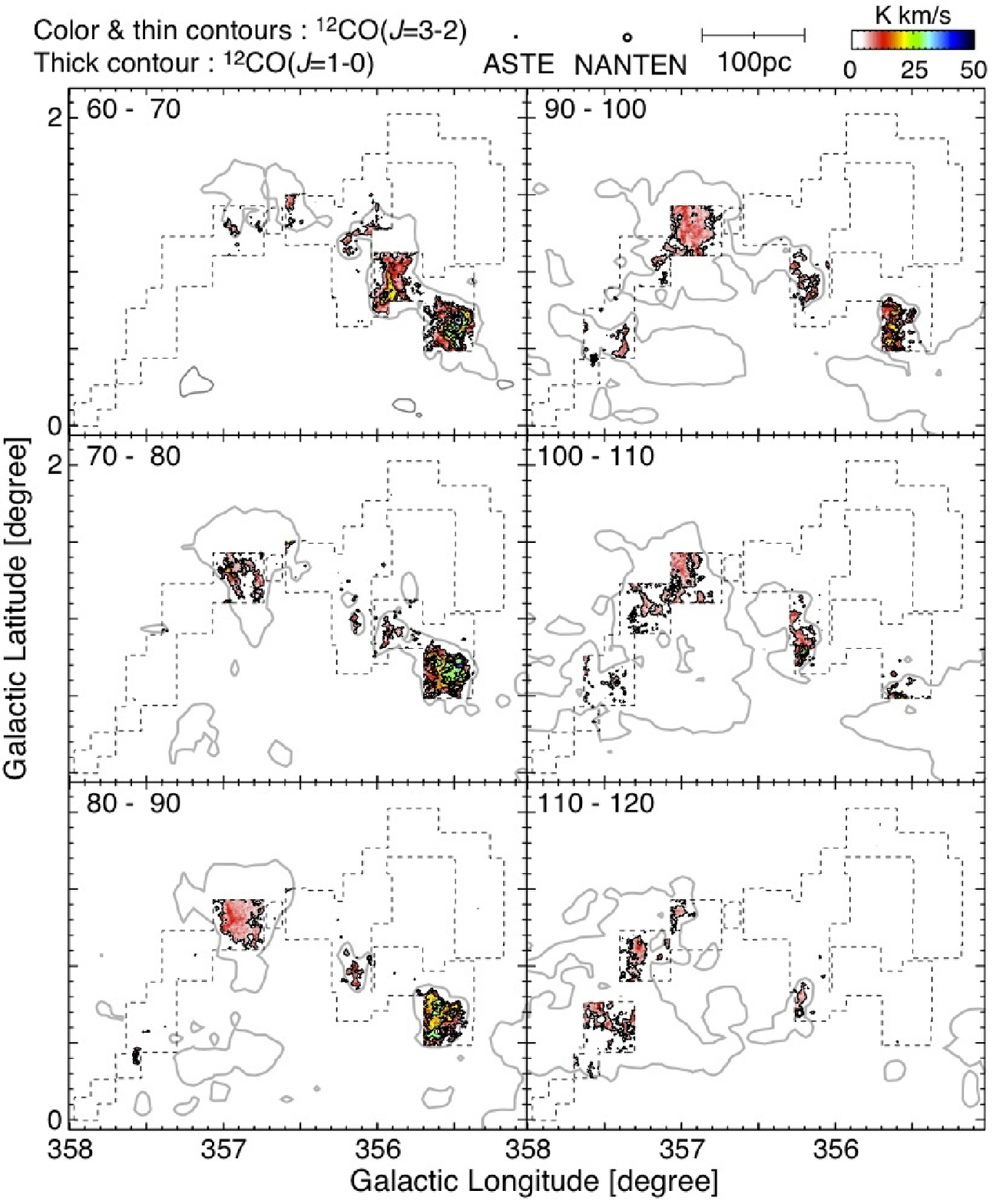}   
  \end{center}
  \contcaption{Continued.}
  \label{fig:channelCO10vsCO32_5}
\end{figure}

\clearpage


\begin{figure}
\renewcommand{\thefigure}{C}
  \begin{center}
    \FigureFile(160mm,160mm){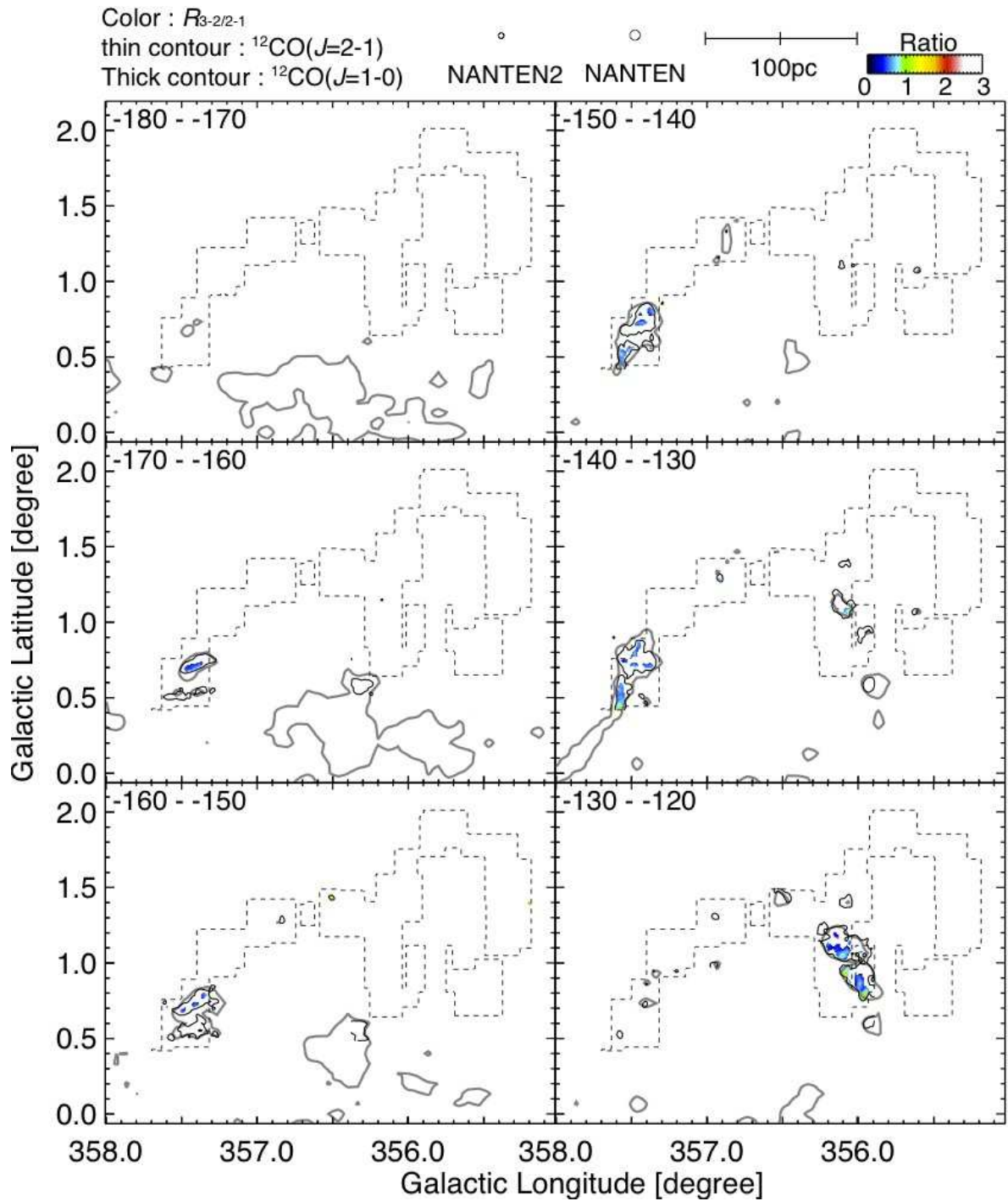}   
  \end{center}
  \caption{Velocity channel distributions of $R_{3-2/2-1}$ integrated every 10 km s$^{-1}$. Thin contours show the $^{12}$CO($J=$2--1) emission and are plotted at 4.7 K km s$^{-1}$. Thick contours show the $^{12}$CO($J=$1--0) emission and are plotted at 7 K km s$^{-1}$. Dashed-lines indicate the analysed regions.}
  \label{fig:channelratioCO21vsCO32_1}
\end{figure}

\begin{figure}
\renewcommand{\thefigure}{C}
  \begin{center}
    \FigureFile(160mm,160mm){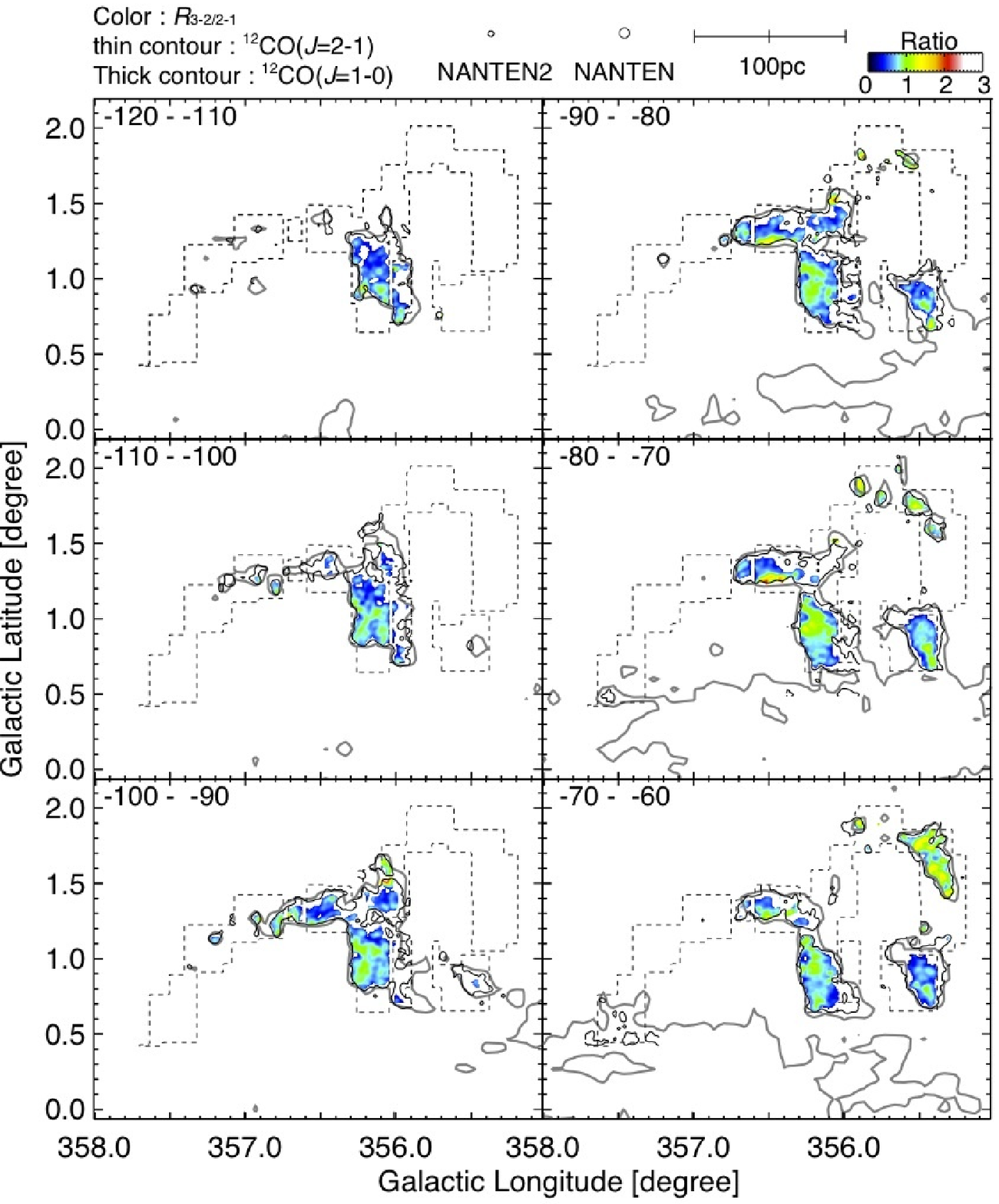}
  \end{center}
  \contcaption{Continued.}
  \label{fig:channelratioCO21vsCO32_2}
\end{figure}

\begin{figure}
\renewcommand{\thefigure}{C}
  \begin{center}
    \FigureFile(160mm,160mm){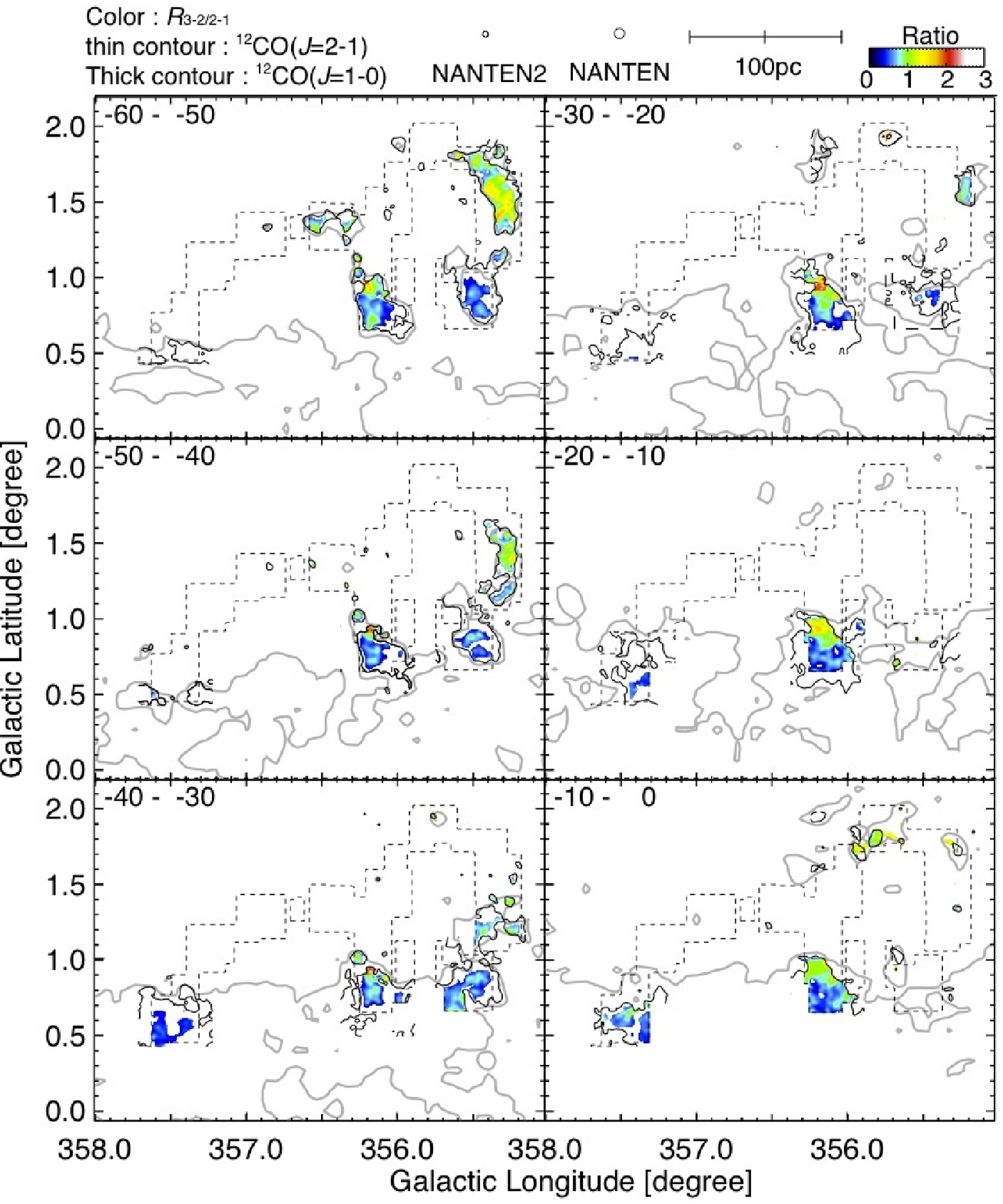}
  \end{center}
  \contcaption{Continued.}
  \label{fig:channelratioCO21vsCO32_3}
\end{figure}

\begin{figure}
\renewcommand{\thefigure}{C}
  \begin{center}
    \FigureFile(160mm,160mm){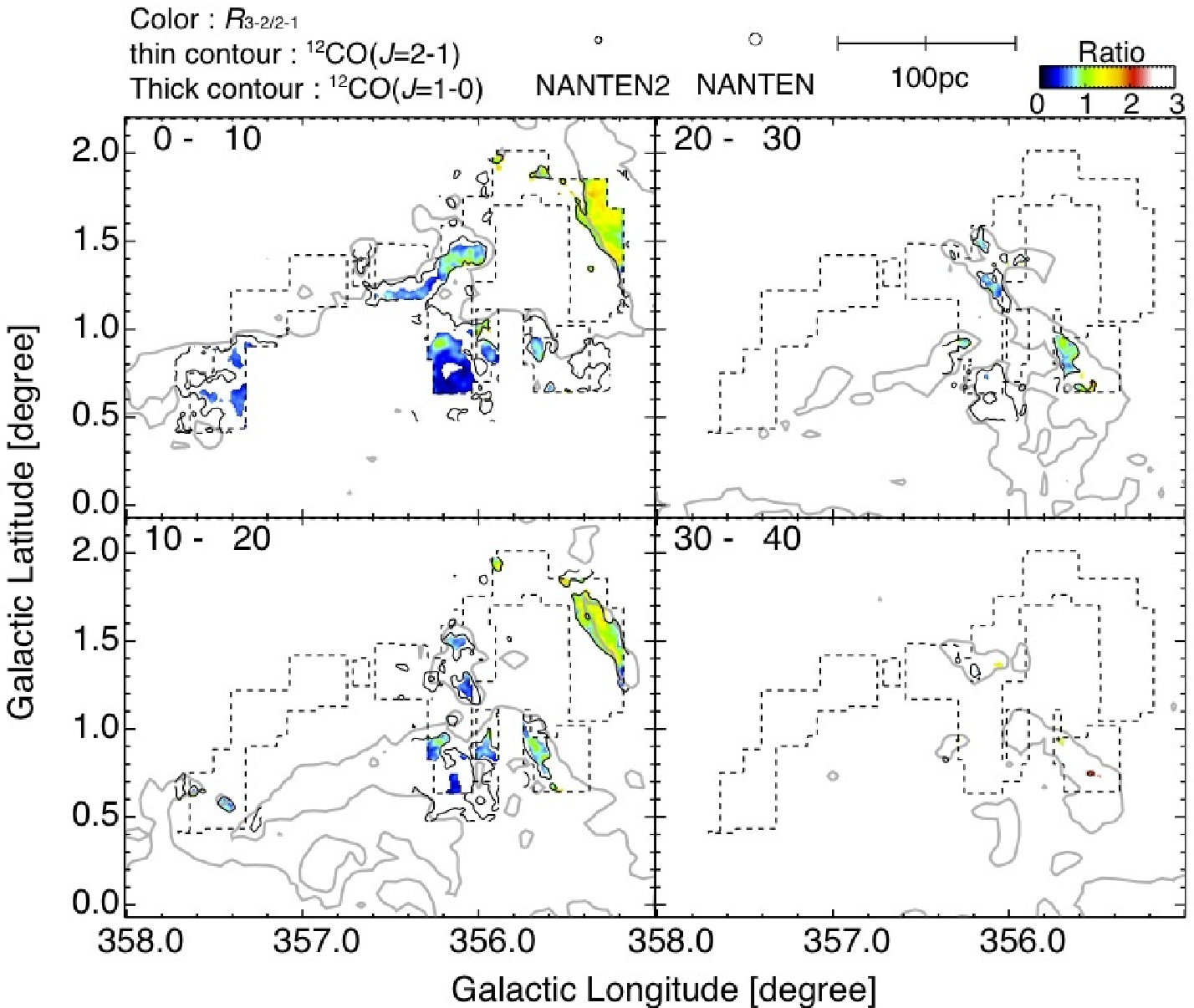}
  \end{center}
  \contcaption{Continued.}
  \label{fig:channelratioCO21vsCO32_3}
\end{figure}

\begin{figure}
\renewcommand{\thefigure}{D}
  \begin{center}
    \FigureFile(120mm,120mm){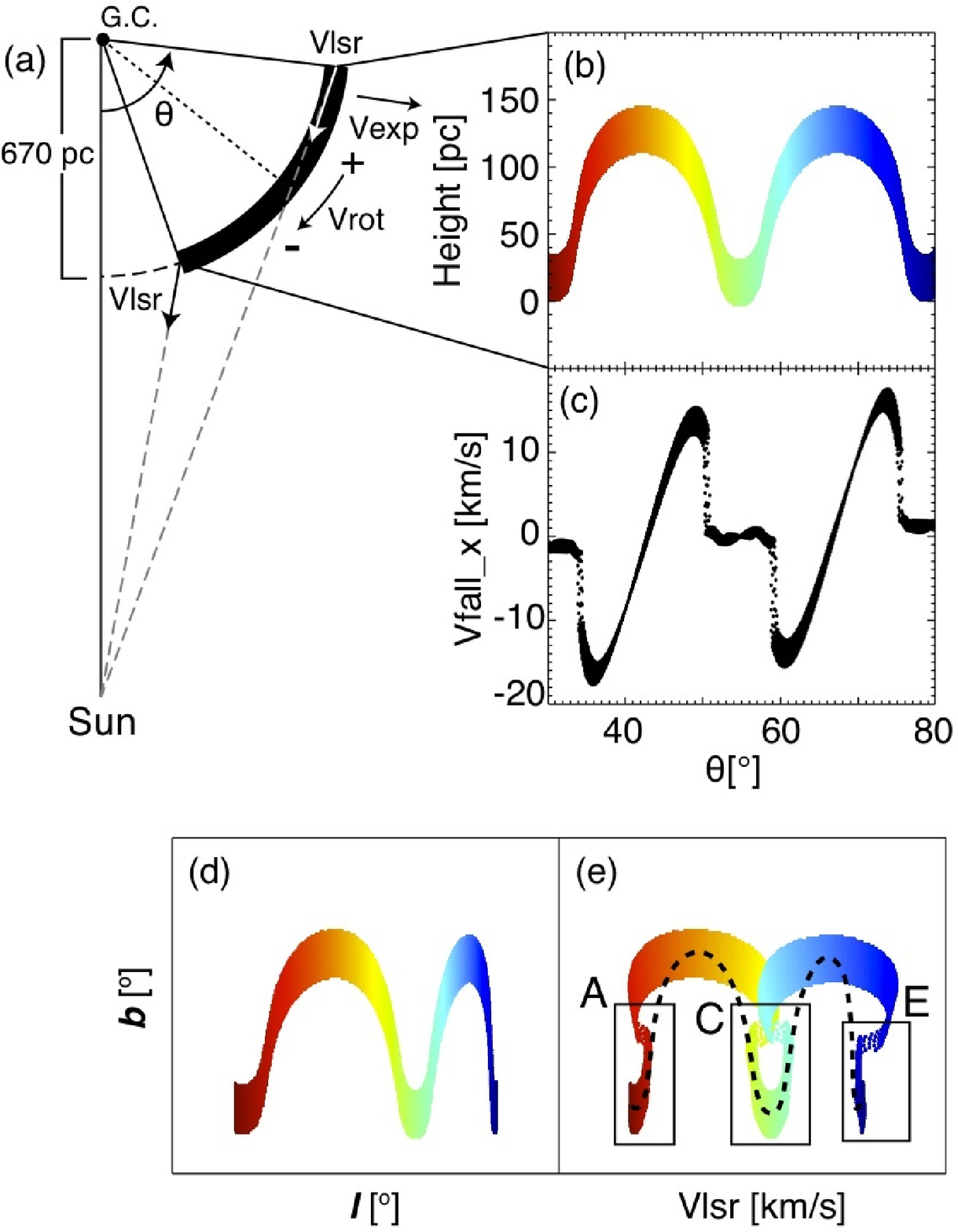} 
  \end{center}
  \caption{(a)Schematic image of the distribution of loops 1 and 2 in a face-on view. (b)Model of two loops on the cylinder coordinate. (c)Down flow speed parallel to the galactic plane on the loops calculated by \citet{kt2009}. (d)Model calculation of the Galactic Longitude-Latitude diagram of two loops estimated with the model. (e) Model calculation of the latitude-velocity diagram of the loops. Dashed-lines indicate the result without down flow speed. Boxes indicate the footpoints of loops 1 and 2 (Regions A, C and E).}
  \label{fig:loopmodel12}
\end{figure}

\begin{figure}
\renewcommand{\thefigure}{E}
  \begin{center}
    \FigureFile(120mm,120mm){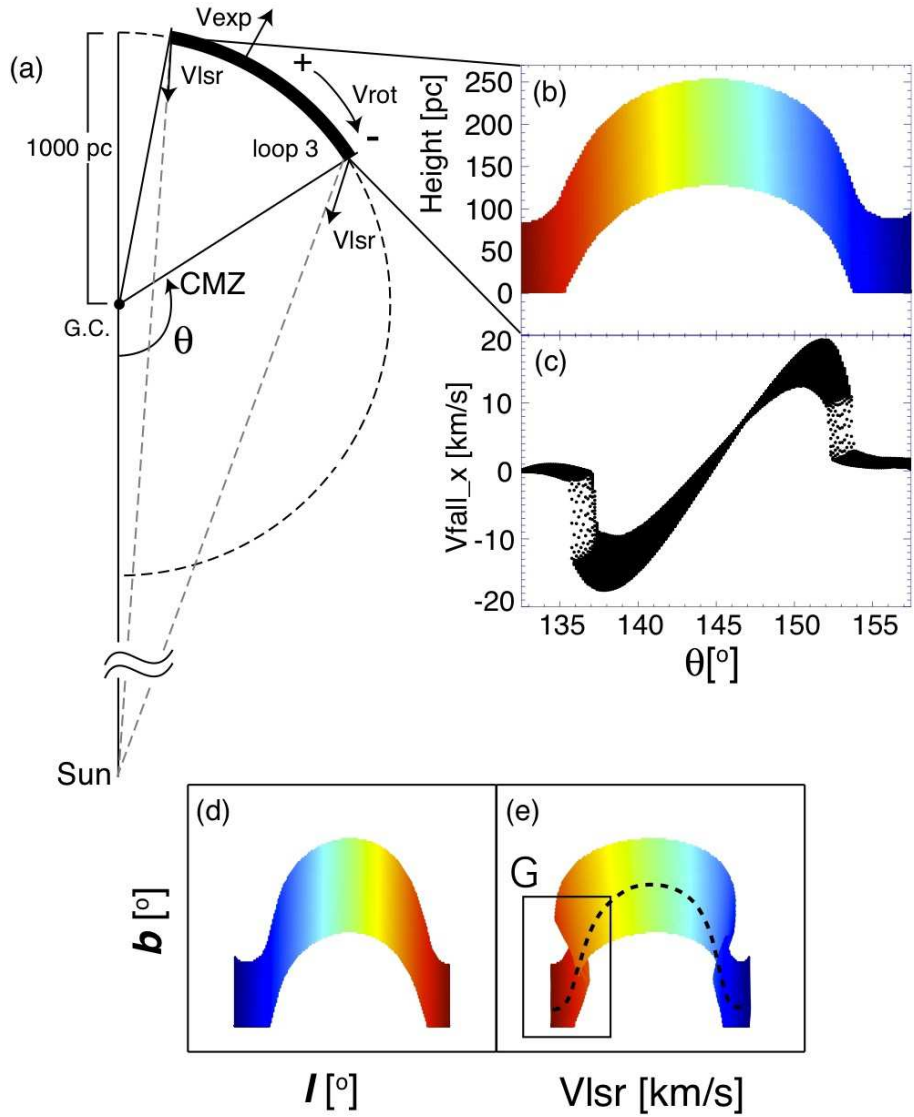} 
  \end{center}
  \caption{(a)Schematic image of the distribution of loop 3 in a face on view. (b)Model of a loop on the cylinder coordinate. (c)Down flow speed parallel to the galactic plane following the loop calculated by \citet{kt2009}. (d)Model calculations of the Galactic longitude-Latitude diagram of the loop. (e)Model calculation of the latitude-velocity diagram of two loops. Dashed- lines indicate the result without down flow speed. Boxes indicate the footpoints of loop 3 (Region G).}
  \label{fig:loopmodel3}
\end{figure}

\end{document}